\documentclass[letter,journal,final,11pt]{IEEEtran}

\usepackage{cite}
\usepackage{graphicx,amsmath,amssymb,latexsym}
\usepackage{epsf,color}
\usepackage[dvips]{epsfig}

\usepackage{amssymb,epsfig,latexsym}

\usepackage{graphicx,amsmath,amssymb,latexsym,epsfig}

\def\cM{{\cal M}}
\def\cV{{\cal V}}
\def\cA{{\cal A}}
\def\cX{{\cal X}}

\def\cI{{\cal I}}

\def\st{{\rm st}}
\def\cF{{\cal F}}

\def\N{\mathbb{N}}
\def\R{\mathbb{R}}

\def\cP{{\cal P}}
\def\cS{{\cal S}}
\def\and{\quad\mbox{and}\quad}

\def\igam{\underline{\gamma}(\lambda)}
\def\sgam{\overline{\gamma}(\lambda)}

\def\PP{{\mathrm P}}

\newtheorem{definition}{Definition}
\newtheorem{prop}{Proposition}
\newtheorem{remark}{Remark}

\newtheorem{theorem}{Theorem}
\newtheorem{lemma}{Lemma}

\def\bp{\noindent{\it Proof.}\ }
\def\ep{\hfill $\Box$}

\begin{document}

\title{Asymptotic stability region of slotted-Aloha}

\author{Charles Bordenave, David McDonald, Alexandre Proutiere
\thanks{C. Bordenave is with CNRS and the Department of Mathematics of Toulouse University, France. D. McDonald is with the Department of Mathematics and Statistics of Ottawa University, Canada. A. Proutiere is with Microsoft Research, Cambridge, UK.
(e-mails: charles.bordenave{\tt @}math.univ-toulouse.fr, dmdsg{\tt @}uottawa.ca, alexandre.proutiere{\tt @}microsoft.com).
}}

\maketitle

\begin{abstract}
We analyze the stability of standard, buffered, slotted-Aloha systems. Specifically, we consider a set of $N$ users, each equipped with an infinite buffer. Packets arrive into user $i$'s buffer according to some stationary ergodic Markovian process of intensity $\lambda_i$. At the beginning of each slot, if user $i$ has packets in its buffer, it attempts to transmit a packet with fixed probability $p_i$ over a shared resource / channel. The transmission is successful only when no other user attempts to use the channel. The stability of such systems has been open since their very first analysis in 1979 by Tsybakov and Mikhailov. In this paper, we propose an approximate stability condition, that is provably exact when the number of users $N$ grows large. We provide theoretical evidence and numerical experiments to explain why the proposed approximate stability condition is extremely accurate even for systems with a restricted number of users (even two or three). We finally extend the results to the case of more efficient CSMA systems.
\end{abstract}


\section{Introduction}

Random multiple access protocols have played a crucial role in the development of both wired and wireless Local Area Networks (LANs), and yet the performance of even the simplest of these protocols, such as slotted-Aloha \cite{A70, roberts72}, is still not clearly understood. These protocols have generated a lot of research interest in the last thirty years, especially recently in attempts to use multi-hop wireless networks (Mesh and AdHoc networks) to provide low-cost high-speed access to the Internet. Random multiple access protocols allow users to share a resource (e.g. a radio channel in wireless LANs) in a distributed manner without exchanging any signaling messages. A crucial question is to determine whether these protocols are efficient and fair, or whether they require significant improvements.

In this paper, we consider non-adaptive protocols, where the transmission probability of a given transmitter is basically fixed. More specifically we analyze the behavior the slotted-Aloha protocol in a buffered system with a fixed number of users receiving packets from independent Markovian processes of pre-defined intensities. We aim at characterizing the stability region of the system. This question has been open since the first stability analysis of Aloha systems in 1979 by Tsybakov and Mikhailov \cite{TM79}, and we will shortly explain why it is so challenging to solve. We propose an approximate stability region and prove that it is exact when the number of users grows large. To accomplish this, we characterize the mean field regime of the system when the number of users is large, explore the stability of this limiting regime, and finally explain how the stability of the mean field regime relates to the ergodicity of systems with a finite number of users. We also show, using both theoretical arguments and numerical results, that our approximate is extremely accurate even for small systems, e.g. with three users (the approximate is actually exact for two users). Our approach can be generalized to other types of non-adaptive random multi-access protocols (e.g., CSMA, Carrier Sense Multiple Access). We present this extension at the end of the paper.

\subsection{Model}\label{sec1}

Consider a communication system where $N$ users share a common resource in a distributed manner using the slotted-Aloha protocol. Specifically, time is slotted, and at the beginning of each slot, should a given user $i$ have a packet to transmit, it attempts to use the resource with probability $p_i$. Let $p=(p_1,\ldots,p_N)$ represent the vector of fixed transmission probabilities.  When two users decide to transmit a packet simultaneously, a collision occurs and the packets of both users have to be retransmitted.

Each user is equipped with an infinite buffer, where it stores the packets in a FIFO manner before there are successfully transmitted. Packets arrive into user $i$'s buffer according to a stationary ergodic process of intensity $\lambda_i$. The arrival processes are independent across users, and are  Markov modulated. More precisely, the packet arrivals for user $i$ can be represented by an ergodic Markov chain $A_i=(A_i(t),t=0,1,\ldots)$ with stationary probability $\pi_i(a)$ of being in state $a$, and with transition kernel $K_i$.  The Markov chains $A_i$ are independent across users and take values
in a finite space $\cA$. If at time slot $t$ $A_i(t)=a$, a new packet arrives into the buffer
of user $i$ with probability $\lambda_{i,a}=\lambda_i \times g_{i,a}$, where the $g_{i,a}$'s are positive real numbers such that $\sum_{a\in {\cal A}}\pi_i(a)g_{i,a}=1$. The average arrival rate of packets per slot at user $i$ is then $\lambda_i$. We use these chains to
represent various classes of packet inter-arrival times. The
simplest example is that of Bernoulli arrivals, i.e., when the
inter-arrivals are geometrically distributed with mean
$1/\lambda_i$: this can be represented by the Markov chain $A_i$
with one state.  We could also represent inter-arrivals that are sums (or random weighted sums) of geometric random variables. In the following we denote by
$\alpha_i=\lambda_i/\sum_j\lambda_j$ the proportion of traffic
generated by user $i$.

Denote by $B_i(t)$ the number of packets in the buffer of user $i$ at the beginning of slot $t$. The state of the system is given by $Z(t)=(A_i(t),B_i(t),i\in\{1,\ldots,N\})$ at time slot $t$. $Z=(Z(t),t=0,1,\ldots)$ is a discrete-time Markov chain. The stability region $\Lambda^N$ is defined as the set of vectors $\lambda=(\lambda_1,\ldots,\lambda_N)$
such that the system is stable, i.e. $Z$ is ergodic, for packet arrival rates $\lambda$. It is important to remark that, a priori, $\Lambda^N$ depends on the transmission probabilities $p$, but also on the types of arrival processes defined by the transition kernels $K=(K_1,\ldots,K_N)$ and the parameters $g=(g_{i,a},i=1,\ldots,N, a\in {\cal A})$. But to keep the notation simple, we use $\Lambda^N$ to denote the stability region.

\subsection{Related work}

The problem of characterizing the stability region $\Lambda^N$ has received a lot of attention in the literature in the three last decades.
First of all  note that when the system is {\it homogeneous} in the sense
 that $\lambda_i/[p_i\prod_{j\neq i}(1-p_i)]$ does not depend on $i$, then one can show as in \cite{BBHP04} that the stability condition is: $\lambda_i<p_i\prod_{j\neq i}(1-p_j)$ for all $i$
regardless of the nature of the arrival process (in this very specific case, all buffers saturate simultaneously at the stability limit).
 For nonhomogeneous systems an exact characterization has been provided in \cite{TM79,SE81,phil85}
under general traffic assumptions but only for $N=2$ users. For two users,
the stability region $\Lambda^2$ is defined by: $\lambda \in \Lambda^2$ if and only if:
$$\hbox{either }\lambda_1<p_1(1-p_2), \lambda_2<p_2(1-\lambda_1/(1-p_2)),$$
$$\hbox{or }\lambda_2<p_2(1-p_1), \lambda_1<p_1(1-\lambda_2/(1-p_1)).$$
The first (resp. second) condition is obtained assuming that at the stability limit, buffer 2 (resp. buffer 1) is saturated. When the number of users is greater than two, the stability region depends not
only on the mean arrival rates $\lambda_i$, but also on the other detailed statistical
properties of the arrival processes. For example, when $N=3$, this is due to the fact that
the stability condition for a particular buffer  depends on the probability that the
two other buffers are empty separately or simultaneously.
These probabilities actually depend on the detailed characteristics of the
arrival processes, see e.g. \cite{Szpan94}. For $N=3$ and Bernoulli arrivals,
the stability region can be characterized \cite{Szpan94}. When the arrivals are not Bernoulli, the system stability region is unknown.

When the number of users $N$ exceeds 3, it becomes impossible to derive explicit stability conditions.
For Bernoulli arrivals, as was shown in \cite{Szpan94}, the stability region $\Lambda^N$ can be  recursively
described as a function of the various stability regions of systems with $N-1$ users, $\Lambda^{N-1}$, and of the probabilities that in these systems,
some buffers are simultaneously empty.
These probabilities are unknown in general, and so is the stability region. The results of \cite{Szpan94} have been recently generalized to more general systems of interacting queues \cite{BJL07}.
The only previous explicit stability condition for arbitrary $N$ is given in \cite{Venkat91}; unfortunately,
to obtain this condition, the author has to assume that the arrival processes of the different
users are correlated, which is unrealistic in practice. Some other authors have proposed bounds on the stability region,
see e.g. \cite{RE88,LE99}. The basic idea behind most of the proposed bounds is to build systems that stochastically dominate (or that are stochastically dominated by) the initial system. For example, a system where one of the buffers is assumed to be always non-empty stochastically dominates the initial system, and hence has a smaller stability region. In \cite{LE06} the reader will find an interesting discussion on the existing techniques to derive bounds of the stability region.

It is worth remarking that often in the literature, researchers have been interested in deriving what we refer to as the {\it capacity region} of Aloha systems. It is defined as the set of vector $\lambda$ such that there exists a vector $p$ of transmission probabilities such that the resulting system is stable. In this paper, we fix the transmission probabilities and investigate the stability region, i.e. the set of $\lambda$ such that the system is stable. In particular, if we succeed in characterizing the stability region for any vector $p$, then we may easily deduce the capacity region.

\subsection{Contributions}

The main contribution of this work is to propose a simple explicit approximate expression of the stability region $\Lambda^N$. This approximate stability region $\hat\Lambda^N$ enjoys the following properties:
\begin{itemize}
\item When the number $N$ of users grows large, the gap between $\hat\Lambda^N$ and the actual stability region $\Lambda^N$ vanishes.
\item Even for small systems, $\hat\Lambda^N$ proves to be very accurate. For $N=2$, one actually has $\hat\Lambda^2=\Lambda^2$; for $N=3$ and any other number of users, the approximate region is very accurate. In fact, for any values of $N$, there exists an infinite number of points where the boundaries of $\Lambda^N$ and of $\hat\Lambda^N$ coincide, which explains the accuracy.
\item $\hat\Lambda^N$ is insensitive, i.e., it depends on the arrival processes through their intensities $\lambda_i$'s only.
\end{itemize}
To prove that $\hat\Lambda^N$ becomes exact when $N$ grows large, we use a mean field analysis of the system, and we show that the stability of the finite system of queues and that of the mean field limiting regime are related (in fact equivalent when $N$ grows large). To our knowledge, this is the first time mean field asymptotics are used  to provide stability conditions of the finite systems.

The paper is organized as follows. In Section II, the approximate stability region is proposed and the main result, i.e. the fact that $\hat\Lambda^N$ tends to $\Lambda^N$ when $N$ is large, is stated in Theorem 1. In Section III, we  present theoretical arguments and numerical experiments to illustrate the accuracy of $\hat\Lambda^N$. Sections IV and V are devoted to the proof of Theorem 1: In Section IV, we present a mean field analysis of the system, and in Section V, we investigate the stability of the system in the limiting mean field regime, and explain why the  stability condition obtained provides an ergodicity condition of the finite system of queues. We generalize our results to non-adaptive CSMA protocols in Section VI, and conclude in Section VII.

\section{Approximate stability region}

\subsection{Approximate stability region $\hat\Lambda^N$}

We now provide an approximate expression of the stability region for
a system with an arbitrary number of users. We prove that this
approximation is exact when the number of users grows large. The
approximate expression is valid for any arrival processes, which
indicates that the stability region becomes insensitive when $N$
grows.

Roughly speaking, the approximate stability region is obtained assuming that the evolutions of the queues of the various users are independent. Let $\partial_j[0,1]^N$ be the set of $\rho\in \R^N_+$ such that $\forall i$, $\rho_i\le 1$, and $\rho_j=1$. The approximate stability region is the region lying below one of $N$ boundaries $\partial_j{\hat{\Lambda}}^N$ defined by:
\begin{align*}
\partial_j\hat\Lambda^N=\bigg\{\lambda: \exists &\rho\in \partial_j[0,1]^N, \forall i,\\
  &\lambda_i=\rho_i p_i\prod_{k\neq i}(1-\rho_kp_k)\bigg\}.
\end{align*}
More precisely, $\hat\Lambda^N$ is the set of positive vectors $\lambda$ such that there
exist $j$ and $\sigma\in \partial_j\hat\Lambda^N$ with $\lambda_i <\sigma_i$ for all $i$. Note that $\hat\Lambda^2=\Lambda^2$, so the proposed approximation
is exact when $N=2$.
\medskip
\begin{remark}[How to compute $\hat\Lambda^N$] Assume that the traffic distribution $\alpha=(\alpha_i, i=1,\ldots,N)$,
is fixed and let us find the maximum total arrival rate $\hat{s}^\star$
such that $\lambda=\hat{s}^\star\alpha$ belongs to the closure of $\hat\Lambda^N$. It can be easily shown that at this maximum,
the user $i^\star$ such that $\rho_{i^\star}=1$ is $i^\star=\arg\max_i \alpha_i(1-p_i)/p_i$. Indeed, since for all $i$, $\alpha_i\hat{s}^\star=\rho_ip_i\prod_{j\neq i}(1-\rho_jp_j)$, then $\alpha_i(1-\rho_ip_i)/(\rho_ip_i)=\alpha_{i^\star}(1-p_{i^\star})/p_{i^\star}$, and
$$
\rho_i<1 \Longleftrightarrow {\alpha_i(1-p_i)\over p_i}<{\alpha_{i^\star}(1-p_{i^\star})\over p_{i^\star}}.
$$
We deduce the maximum arrival rate:
$$
\hat{s}^\star={p_{i^\star}\over \alpha_{i^\star}}\prod_{i\neq i^\star}\left({\alpha_{i^\star}(1-p_{i^\star})\over \alpha_ip_{i^\star}+\alpha_{i^\star}(1-p_{i^\star})}\right).
$$
\end{remark}

\subsection{Main result}

Our main result states that the actual stability region $\Lambda^N$ is very close to the proposed approximation $\hat\Lambda^N$ when $N$ is large. To formalize this, we introduce a sequence of systems indexed by $N$, i.e., the arrival rates are $\lambda^N=(\lambda_1^N,\ldots,\lambda_N^N)$, the transmission probabilities are $p^N=(p_1^N,\ldots,p_N^N)$, and the Markov chains modulating the arrival processes are $A_1^N,\ldots,A_N^N$.

In order for  a system with $N$ users to give reasonable bandwidth to each user the duration of a time slot must be of order $1/N$ seconds, i.e., we suppose $\tau$ seconds will represent $t=\lfloor N\tau\rfloor$ time slots. We assume that users can be categorized among a finite set ${\cal V}$ of $V=\vert\cV\vert$ classes. Further we assume the proportion of users $i$ in class $v$ tends to $\beta_v$ when $N\to\infty$. The class of a user characterizes its transmission probability and the packet arrival process in its buffer. The transmission probability of user $i$ of class $v$ is $p_i^N=p_v/N$. We assume that for all $N$, $\sum_ip_i^N \le 1$. This assumption is made so as to keep the approximation expression of the stability region simple. In Section V, we explain how to extend the assumption, and give ways to remove it. Note that, as Kleinrock already noticed \cite{kleinrock}, the assumption is needed to guarantee a certain efficiency of the system.

In order that the system not be overloaded, we assume that the mean packet arrival rate of user $i$ of class $v$ is $\lambda_i^N=\lambda_v/N$. The class of a user also defines the Markov chain modulating its arrival process. If user $i$ is of class $v$, this Markov chain $A_i^N$ is assumed to be in stationary regime with probability $\pi_v(a)$ of being in state $a$, in which case the probability that a packet arrives in its buffer is $\lambda_{i,a}^N=\lambda_v\times g_{v,a}/N$, where $g_{v,a}$ are positive real numbers such that $\sum_{a\in \cA}\pi_v(a)g_{v,a}=1$. We assume that the Markov chains $A_1^N, \ldots, A^N_N$ are independent. Denote by $K_i^N$ the transition kernel of $A_i^N$. The Markov chains $A_i^N$ may be fast or slow.  In particular the arrival rate may change on the order of time slots if packets are generated by a HTTP connection since the transmission rate is changed dynamically by HTTP. On the other hand the state describing a VoIP connection would evolve  on the scale of seconds as the speaker alternates between silent and speech periods. To represent these two scenarios, we introduce $V$ continuous-time Markov processes $A_v=(A_v(\tau),\tau\in \mathbb{R}_+)$, $v\in {\cal V}$, taking values in ${\cal A}$ with jump rate kernel $K_v$ (expressed in transitions per second), and we assume that the corresponding Markov chains modulating the arrivals of the various users are as follows.

\medskip
{\it Sequence of type 1 - Fast modulated arrivals.} In this case, for all $t=0,1,\ldots$, the law of $A_i^N(t)$ is equal to that of $A_v(t)=A_v(\lfloor N\tau\rfloor)$ (recall that $\tau$ seconds roughly represent $N\tau$ slots).  In other words,  the modulating Markov chain
changes state on the scale of time slots; i.e. $N$ times faster than the speed at which the user evolves  (e.g. the speed at which the user attempts to transmit a packet). In this case $K_i^N(a,a')=P(A_v(1)=a'|A_v(0)=a)$.

\medskip
{\it Sequence of type 2 - Slowly modulated arrivals.} In this case, for all $t=0,1,\ldots$,  the law of $A_i^N(t)$ is equal to that of $A_v(t/N)=A_v(\lfloor N\tau\rfloor/N)$; i.e. the speed of the modulating Markov chains  is proportional to the speed of at which the user evolves. In this case $K_i^N(a,a')=P(A_v(1/N)=a'|A_v(0)=a)\approx K_v(a,a')/N$

\medskip
We denote by $\Lambda^N$ the stability region of the system indexed by $N$. As explained earlier, the stability region depends on the transmission probabilities and on the type of arrival processes. The following theorem compares the stability region $\Lambda^N$ with our proposed approximation $\hat \Lambda^N$ as $N$ gets large. The theorem is valid for both types of sequences of systems, 1 and 2. Since $\hat\Lambda^N$ does not depend on the kernels $K_i^N$, the theorem indicates that when $N$ is large, the stability region depends on the arrival processes through the mean arrival rates only. Define $1^N:=(1/N,\ldots,1/N)$.

\medskip
\begin{theorem}\label{th:stab1}
For all $\epsilon>0$ small enough, there exists $N_\epsilon$ such that for all $N>N_\epsilon$:\\
(a) if $\lambda^N + \epsilon \cdot 1^N \in \hat
 \Lambda^N$, then $\lambda^N\in\Lambda^N$;\\
(b) if $\lambda^N - \epsilon \cdot 1^N \not\in \hat \Lambda^N$,
then $\lambda^N\not\in\Lambda^N$.
\end{theorem}

\medskip
Theorem \ref{th:stab1} is proven in Sections IV and V.  The main steps of the proof are as follows.\\
(1) The evolution of the system when $N$ grows large is characterized: it is shown that with an appropriate scaling in time, the evolution of the distributions of the various queues is the solution of a deterministic dynamical system. This result is obtained using mean field asymptotic techniques, as presented in Section IV.A. These techniques typically provide an approximate description of the evolution of the system over finite time horizons. Here we wish to study the ergodicity of slotted-Aloha systems, which basically relates to the system dynamics over an infinite horizon of time. Hence, classical mean field asymptotic results will be necessary but not sufficient to prove Theorem 1. The main technical contribution of this paper is to explain how mean field asymptotics can be used to infer the ergodicity of the finite systems, and this is what is done in steps (2) and (3).\\
(2) We provide sufficient and necessary conditions for the global stability of the dynamical system describing the evolution of the system in the mean field regime. \\
(3) Finally, it is shown that the ergodicity of the initial system of $N$ queues is equivalent to the stability of the mean field dynamical system when $N$ grows large.

\medskip
\begin{remark}[Capacity of Aloha systems]
A consequence of the above result is that when $N$ grows large,
and whatever the arrival processes considered, the set traffic
intensities $\lambda$ such that there exist transmission
probabilities $p$ stabilizing the system is the set $\cM^N$
with boundary $\partial \cM^N$:
\begin{align*}
\partial \cM^N=\bigg\{\lambda:& \exists p_1,\ldots,p_n\in (0,1): \\
&\forall i,  \lambda_i=p_i\prod_{j\neq i}(1-p_j)\bigg\}.
\end{align*}
This result has been conjectured by Tsybakov and Mikhailov in \cite{TM79}. It has been proved in \cite{Venkat91}, but under the assumption that the arrival processes of the various users are correlated. The authors of \cite{LE06} have introduced the so-called {\it sensitivity monotonicity} conjecture under which they could also prove the result. Theorem 1 says that when the number of users is large enough, the sensitivity monotonicity conjecture is not needed.
\end{remark}

\medskip
\begin{remark}[Shannon capacity of Aloha systems]
It is worth noting that $\cM^N$ coincides with the Shannon capacity region of the multi-user collision channel derived in \cite{MM85, H84}. Theorem 1 shows that the capacity region and the Shannon capacity region are equivalent when the number of users grows large. In \cite{SE07} the reader will find  a more detailed discussion on the comparison of these two regions in communication systems.
\end{remark}

\section{Accuracy of $\hat\Lambda^N$}

How far is the approximate region $\hat\Lambda^N$ from the actual
stability region? Theorem \ref{th:stab1} says that the gap tends to
0 when the number $N$ of users grows large. But even for small $N$,
$\hat\Lambda^N$ is  quite an accurate approximation as illustrated
in the numerical examples provided later. Why is the approximate region so accurate?

\subsection{$k$-Homogeneous systems}

This accuracy can be explained by  remarking that the boundaries of the regions
$\Lambda^N$ and $\hat\Lambda^N$ coincide in many scenarios. Remember that $\hat\Lambda^N$ can be interpreted as the
stability region one would get if the evolutions of the different
buffers were independent. As a consequence, it provides the exact
stability condition for scenarios where, in the stability limit, the
buffers become independent.

\medskip
\begin{definition}[$k$-homogeneous directions] A direction (a vector with unit $L_1$-norm) $\alpha\in \R^N_+$ is $k$-homogeneous for the system considered if there exists a permutation $\sigma$ of $\{1,\ldots,N\}$ such that, for all $i=1,\ldots,k$, $\alpha_{\sigma(i)}(1-p_{\sigma(i)})/p_{\sigma(i)}$ does not depend on $i$.
\end{definition}

\begin{figure}[htp]
\begin{center}
\scalebox{0.41}{\input{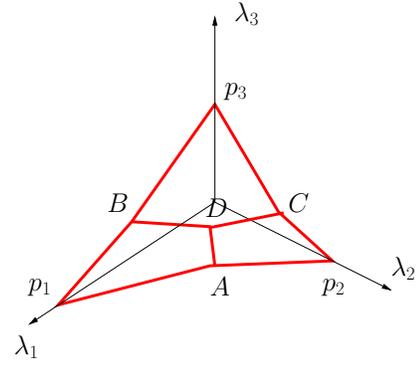}}
\end{center}
\caption{Curves where $\partial\hat\Lambda^3$ and $\partial\Lambda^3$ coincide. $A=(p_1(1-p_2),p_2(1-p_1),0)$; $B=(p_1(1-p_3),0,p_3(1-p_1))$; $C=(0,p_2(1-p_3),p_3(1-p_2))$; $D=(p_1(1-p_2(1-p_3),p_2(1-p_1(1-p_3),p_3(1-p_1)(1-p_2))$.}
\label{fig:ex3}
\end{figure}

\medskip
In the following, without loss of generality, when a direction is $k$-homogeneous, the corresponding permutation is given by $\sigma(i)=i$ for all $i$. The following proposition, proved in appendix, formalizes the fact that the boundaries of $\hat\Lambda^N$ and $\Lambda^N$ coincide on a set of curves corresponding to particular directions, $k$-homogeneous directions. In case $N=3$, Figure \ref{fig:ex3} gives a schematic illustration of these curves.

\medskip
\begin{prop}
Assume that $\lambda=s\times \alpha$, where $\alpha$ is a \\$k$-homogeneous direction for the system considered. Define $s^\star=\sup\{s\ge 0: s\alpha \in \Lambda^N\}$ and similarly $\hat{s}^\star$. Then if $1_{k+1\le N}\alpha_{k+1}(1-p_{k+1})/p_{k+1}\le  \alpha_{1}(1-p_{1})/p_1$ and $\alpha_l=0$ for $l>k+1$, then:
$$
s^\star=\hat{s}^\star={\prod_{i=1}^k(1-p_i)\over {1-p_1\over p_1}\alpha_1+\alpha_{k+1}}.
$$
\end{prop}

\medskip

\subsection{Numerical examples}
\begin{figure*}[ht]
\centering
\includegraphics[width=0.66\columnwidth]{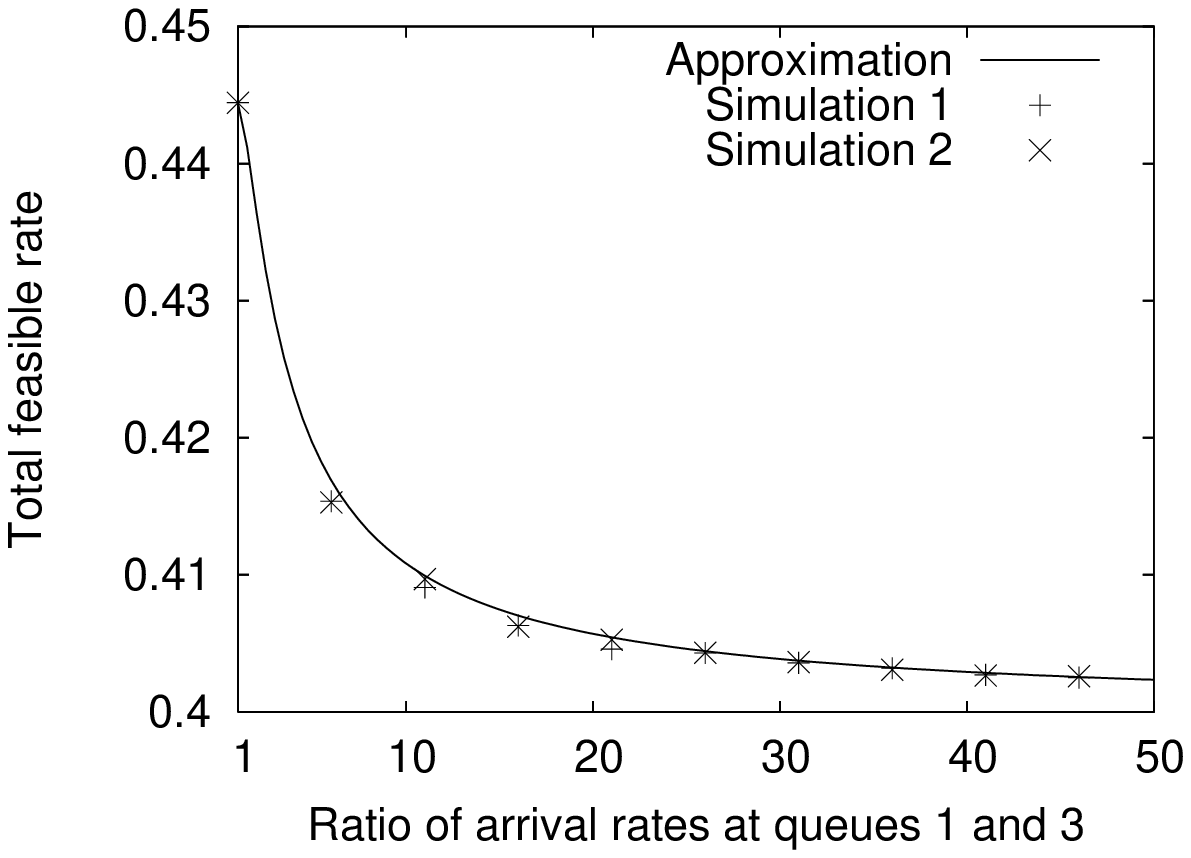}
\includegraphics[width=0.66\columnwidth]{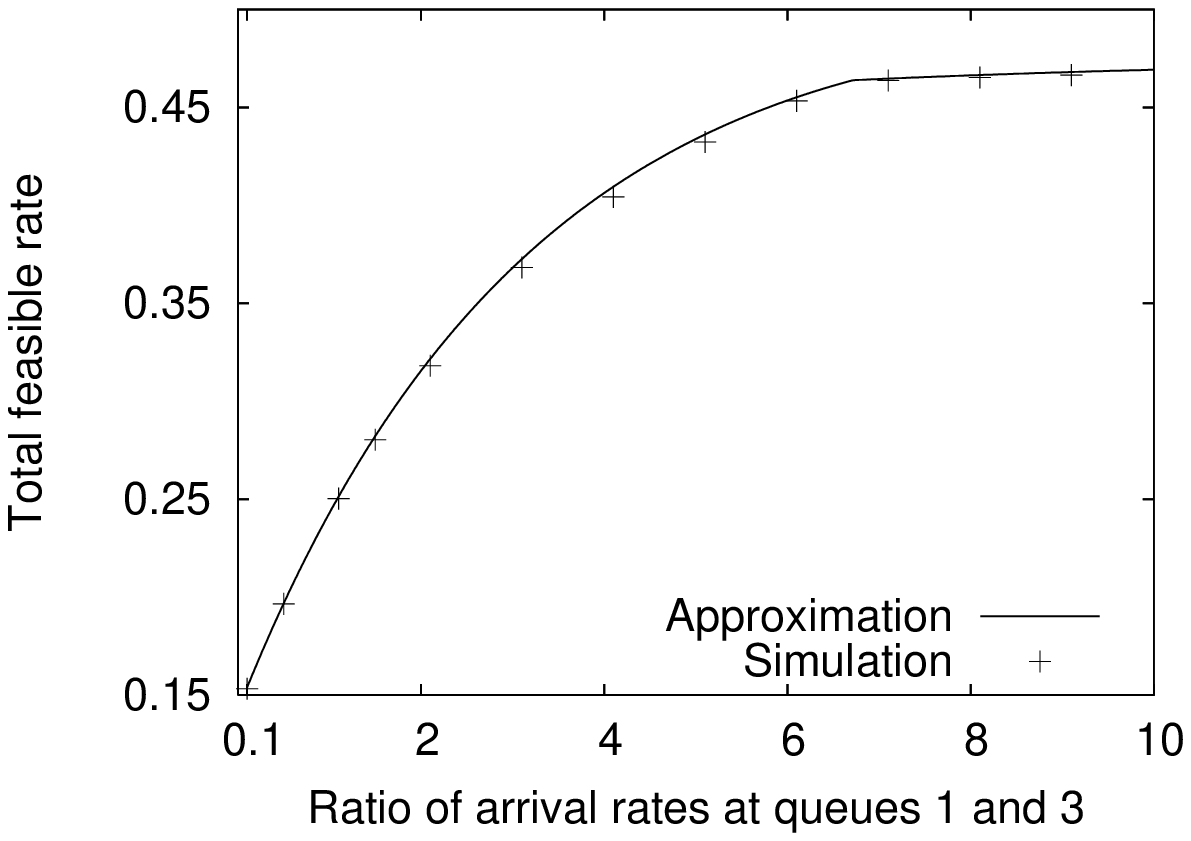}
\includegraphics[width=0.66\columnwidth]{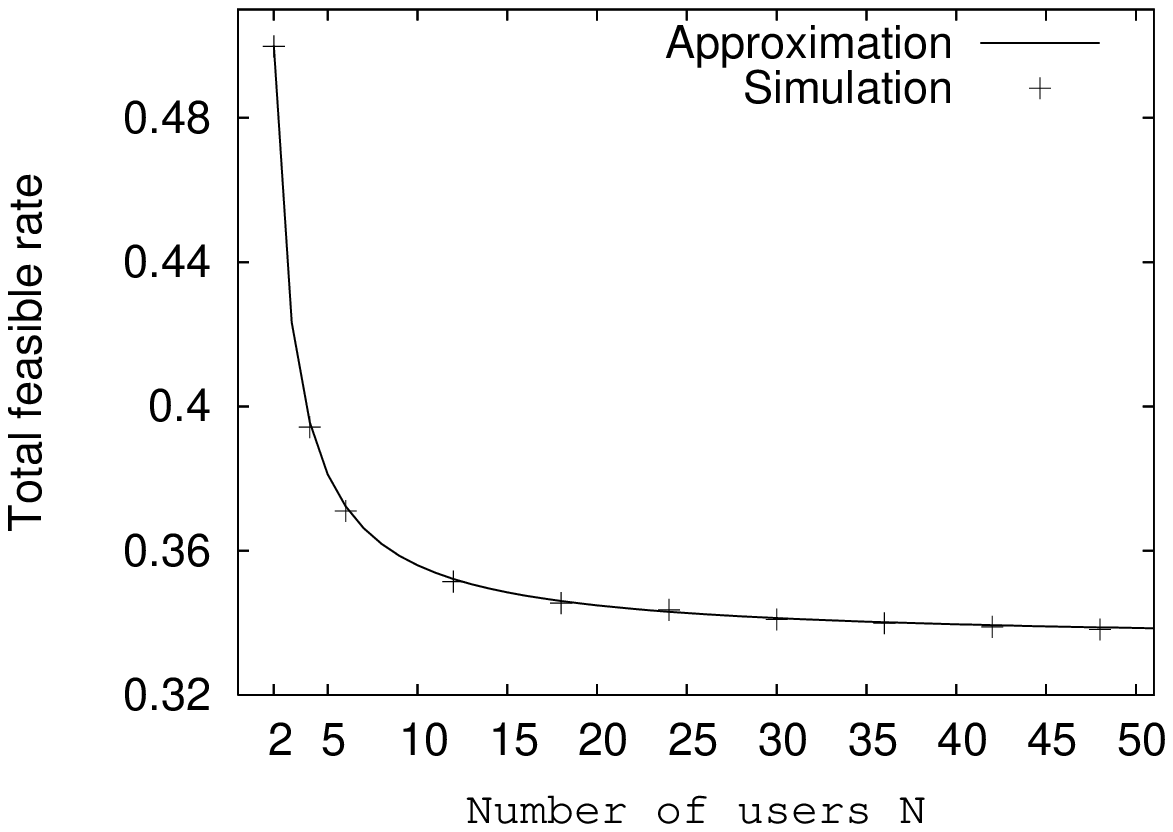}
\caption{Maximum total rate compatible with stability - Left: Example 1, $N=3$, $(p_1,p_2,p_3)=(0.6,0.3,0.1)$ - Center: Example 2, $N=3$, $(p_1,p_2,p_3)=(0.6,0.3,0.1)$ - Right: Example 3, $N$ varies, linearly decreasing traffic distribution $\beta_i$.}
\label{fig:comp}
\end{figure*}

We now illustrate the accuracy of $\hat\Lambda^N$ using numerical experiments.\\
\underline{Example 1:} First, we consider the case of $N=3$ sources,
each transmitting with probability 1/3. We vary the relative values
of the arrival rates at the various queues: $\lambda_1=\lambda$,
$\lambda_2=\lambda\times {(1+1/x)\over 2}$ and
$\lambda_3=\lambda/x$. We vary $x$ from 1 to 50. It can be shown
that the approximate stability condition is
$$
\sum_{i=1}^3\lambda_i < \hat{s}^\star={4x(x+1)\over (2x+1)(5x+1)}.
$$
In Figure \ref{fig:comp} (left), we compare this limit to the actual
stability limit found by simulation with Bernoulli arrivals (Simulation 1) and hyper-geometric arrivals (Simulation 2). In the latter case, the inter-arrivals for each user $i$ are i.i.d., and an inter-arrival is a geometric random variable with parameter $a\lambda_i$ with probability 1/2, and $(1-a)\lambda_i$ with probability 1/2. This increases the variance of inter-arrivals (when $a$ is small the variance scales as $1/a$). In the numerical experiment, we chose $a=1/5$. Remark that the stability region is roughly insensitive to the distribution of inter-arrivals. This insensitivity has been also observed in the other examples presented in this section. The simulation results have been obtained running the system for about $10^7$ packet arrivals. Note finally that the arrival rates are chosen so that the system is not $k$-homogeneous.

\noindent \underline{Example 2:} We make a similar numerical
experiment when $p_1,p_2,p_3$ are equal to 0.6, 0.3, 0.1
respectively. The arrival rates at the three queues are as in
Example 1. We vary $x$ from 0.1 and 10.  For $x<x_0=47/7$, at the
boundary of $\hat{\Lambda}^3$, queue 3 is saturated ($\rho_3=1$);
whereas for $x\ge x_0$, queue 2 is saturated ($\rho_2=1$). The
approximate stability condition is:
$$
\sum_{i=1}^3\lambda_i<\hat{s}^\star=\left\{
\begin{array}{ll}
{7.2x(x+1)\over (7x+3)(2x+3)},& \hbox{if }x<47/7,\\
{44.1(x+1)^2\over (13x+7)(7x+13)},& \hbox{if }x\ge 47/7.
\end{array}
\right.
$$
Figure \ref{fig:comp} (center) illustrates the accuracy of
$\hat\Lambda^3$.

\noindent \underline{Example 3:} Finally, we illustrate the accuracy
of $\hat\Lambda^N$ when the number of users $N$ grows. Each user is
assumed to transmit with probability $1/N$ and the traffic
distribution is such that $\alpha_1>\alpha_i$ for all $i\ge 2$.
Hence, again the system is not $k$-homogeneous. One can easily show
that in this direction, the approximate stability condition is:
$$
\sum_{i=1}^N\lambda_i< \hat{s}^\star={1\over N\alpha_1}\prod_{i=2}^N\left(1-{\alpha_i\over \alpha_i+(N-1)\alpha_1}\right).
$$
In  Figure \ref{fig:comp} (right), we compare the boundary of $\hat\Lambda^N$ with that of $\Lambda^N$
when the distribution $\alpha_i$ is linearly decreasing with $i$.
Again as expected, $\hat{s}^\star$ provides an excellent approximation of the saturation level in the actual system.

The two next sections are devoted to the proof of Theorem 1. In section IV, we provide a classical mean field analysis of the system, and in Section V, we show how the stability in the limiting mean field regime translates into the ergodicity of the initial finite systems.

\section{Mean field asymptotics}

In this section, we first present a generic mean field analysis of a system of interacting particles and then apply the  results obtained to slotted-Aloha systems. Let us first give some notations.

\medskip
\noindent
{\it Notations.} Let ${\cal Y}$ be a complete separable metric space, ${\cal P}({\cal Y})$ denotes the space of probability measures on ${\cal Y}$. ${\cal L}(X)$ denotes the distribution of the ${\cal Y}$-valued random variable $X$. Let $D(\mathbb{R}^+,{\cal Y})$ be the set of right-continuous functions with left-handed limits, endowed with the Skorohod topology associated with the metric $d^0_{\infty}$, see \cite{billingsley} p 168. With this metric, $D(\mathbb{R}^+,{\cal Y})$ is complete and separable. For two probability measures $\alpha,\beta$, we denote by $\| \alpha - \beta\|$ their distance in total variation.

\subsection{A generic particle system and its mean field limit}

Consider a system of $N$ particles evolving in a state space $\cV\times \cX$ at discrete time slots $t\in\mathbb N$. $\cV$ is a finite set, and $\cX$ is at most countable. At time $t$, the state of particle $i$ is $X_i ^N (t)=(v_i^N,Y_i^N(t)) \in \cV\times\cX$. The first component $v_i^N$ of $X_i^N(t)$ is fixed, and is used to represent the {\it class} of a particle as explained below.
$Y^N_i(t)$ represents the state of particle $i$ at time $t$. The state of the system at time $t$ can be described by the empirical measure $\nu^N(t) = \frac 1 N \sum_{i=1} ^ N \delta_{X^N_i(t)}
\in \cP(\cX)$.

Each particle $i$ is attached to an {\it individual environment} whose state $A_i^N(t)$ at time slot $t$ belongs to a finite space ${\cal A}$. $A_i^N=(A_i^N(t),t=0,1,\ldots)$ is a Markov chain with kernel $K_i$ independent of $N$. Particles of the same class share the same kernel: $i,i'\in v$ implies $K_i=K_{i'}$. The Markov chains $A_i^N$ are independent across particles, and are assumed to be in stationary regime at time 0. Let $\pi_v$ be the stationary distribution of the individual environment of a class-$v$ particle.
%whose evolution depends on the class $v\in {\cal V}$ of the particle but is independent of everything %else. and  where $v\in \cV$ denotes the class of the particle, i.e., $X_i^N(t)=(v, Y_i^N(t))$. We assume %that for all $v$, $K_v^N$ converges to $K_v$ when $N\to\infty$ and that $A_i^N(t)$ is in stationary regime %for all $t$.

{\it Evolution of the particles.} We represent the possible
transitions for a particle by a finite set $\cS$ of mappings from
$\cX$ to $\cX$. A $s$-transition for a particle in state $x=(v,y)$ leads
this particle to the state $s(x)=(v,s(y))$. In each time slot the state of a
particle has a transition with probability $1/N$ independently of
everything else. If a transition occurs for a particle whose individual environment is in state $a\in{\cal A}$, this transition is a $s$-transition with probability $F^N _s (x,\nu,a)$, where $x$, $a$, and
$\nu$ denote the state of the particle,  the
empirical measure before the transition and the state of its individual environment respectively. Hence, in this state, a $s$-transition occurs with
probability:
\begin{eqnarray}\label{jumps}
\frac 1 N F^N _s (x,\nu,a).
\end{eqnarray}
with $\sum_{s\in\cS} F^N_s(x,\nu,a) = 1$ for all $(x,\nu,a)$.

Note that we do not completely specify the transition kernel of the Markov chain $((X_i^N(t), A_i^N(t), i=1,\ldots,N),t=0,1,\ldots)$. All what we require is that each particle has a transition with probability $1/N$ independently of the other particles. However given that transitions occur for two (or more) particles, these transitions can be arbitrarily correlated (but with marginals given by (\ref{jumps})). Note also that the chains $A_i^N$ evolve quickly compared to $X_i^N$.

We make the following assumptions on the transition probabilities $F^N_s$.\\
A1. Uniform convergence of $F^N_s$ to $F_s$:\\
$$
\sup_{(x,\alpha,a)\in {\cal X}\times {\cal P}({\cal X})\times {\cal A}}\sum_{s\in {\cal S}}\vert F^N_s(x,\alpha,a)-F_s(x,\alpha,a)\vert \stackrel{N\to\infty}{\longrightarrow}0.
$$
A2. The functions $F_s$ are uniformly Lipschitz: for all $\alpha,\beta\in {\cal P}({\cal X})$,
$$
\sup_{(x,a)\in {\cal X}\times{\cal A}} \sum_{s\in {\cal S}}\vert F_s(x,\alpha,a)-F_s(x,\beta,a)\vert \le \| \alpha-\beta\|.
$$

In what follows, we characterize the evolution of
the system when the number of particles grows. According to
(\ref{jumps}), as $N\to\infty$, the evolution of $X^N_i(t)$ slows down (where $t$ is measured in slots).
Hence to derive a limiting behavior we define: $q_i ^ N (\tau) = X^N_i(\lfloor N\tau\rfloor)$ where $\tau$ is measured in seconds.
When $N\to\infty$, the environment processes evolve rapidly, and the particles see an average of the environments. We define the average transition rates for a particle in state $x=(v,y)$ by
\begin{equation}
\label{eq:Fbarre}
\overline F_s
(x,\alpha)= \sum_{a\in \cA} F_s (x,\alpha,a)\pi_v(a).
\end{equation}

\subsubsection{Transient regimes}

\medskip
\begin{theorem}\label{th:mainApp}
Suppose that the initial values
$q_i^N(0)$, $i=1,\ldots,N$, are i.i.d. and such that their
empirical measure $\mu^N_0$ converges in distribution to a
deterministic limit $Q_0\in \cP(\cX)$. Then under Assumptions A1 and A2, there exists a probability
measure $Q$ on $D(\R^+,\cX)$ such that for all finite set $\cI$ of $I$ particles:
$$
\lim_{N\to\infty} {\cal L}(q_i^N(.),i\in \cI)=Q^{\otimes I},\quad \hbox{weakly.}
$$
\end{theorem}

\medskip
In the above theorem $q_i^N(\cdot)$ denotes the trajectory of particle $i$, which is a random variable taking values in $D(\mathbb{R}^+,{\cal X})$. The result is then stronger than having the weak convergence of the distribution of $q_i^N(\tau)$ in ${\cal P}({\cal X})$ for any $\tau$. For instance, it allows us to get information about the time spent by a particle in a given state during time interval $[0,1]$.

The theorem states that the trajectories of the particles becomes independent when the number of particles grows large. The independence allows us to derive an explicit expression for the system state evolution.\\ Define $Q^n(\tau) = Q(\tau)(\{x_n\})$ where $\cX = \{ x_n, n\in\N\}$. $Q^n(\tau)$ is the limiting (when $N\to\infty$) probability that a particle is in state $x_n$ at time $\tau$, $Q^n(\tau)=\lim_{N\to\infty}Pr[q_i^N(\tau)=x_n]$.
\medskip
\begin{theorem}\label{theoex}
For all time $\tau>0$, for all $n\in \N$,
\begin{eqnarray}\label{eqdiff1}
{d{Q^n}\over d\tau}& =& \sum_{s \in \cS} \sum_{m :
s(x_m) = x_n } Q^m (\tau) \overline{F}_s ( x_m, Q(\tau))\nonumber\\
&&  -\sum_{s \in \cS} Q^n (\tau)\overline{F}_s ( x_n, Q(\tau)).
\end{eqnarray}
\end{theorem}

\medskip
The differential equations (\ref{eqdiff1}) have a natural simple
interpretation:
$$\sum_{s \in \cS} \sum_{m : s(x_m) = x_n } Q^m (\tau) \overline F_s (x_m, Q(\tau))$$ is the total mean incoming flow of particles to state $x_n$, whereas
$$\sum_{s \in \cS} Q^n (\tau) \overline F_s ( x_n, Q(\tau))$$
is the mean outgoing flow from $x_n$.

\medskip
\subsubsection{Stationary regime}

Theorems 2 and 3 characterize the limiting system evolution on all compacts in time. Hence, they do not say anything about the long-term behavior of the system. Here we will assume that the finite particle systems are ergodic and describe the mean field regime of the systems in equilibrium. To do so, we need two additional assumptions:\\
A3. For all $N$ large enough, the Markov chain $((X_i^N(t),A_i^N(t),i=1,\ldots,N), t=0,1,\ldots)$ is positive recurrent. The set of the stationary distributions $\pi_{st}^N$ of the systems with $N$ particles is tight.\\
A4. The dynamical system (\ref{eqdiff1}) is globally
stable: there exists a measure $Q_{st} = (Q^n_{st}) \in \cP(\cX)$
satisfying for all $n$:
\begin{equation}\label{eq:balance}
\sum_{s \in \cS}  \sum_{m : s(x_m) = x_n } Q^m _ {st}\overline F_s (
x_m, Q_{st}) =  Q^n _ {st}\sum_{s \in \cS} \overline F_s ( x_n,
Q_{st}),
\end{equation}
and such that for all $Q \in \cP(D(\R^+,\cX))$ satisfying
(\ref{eqdiff1}), for all $n$, $\lim_{\tau\to +\infty} Q^n (\tau)=
Q_{st}^n$.
Then the asymptotic independence of the particles  also holds in the
stationary regime, and $Q_{st}$ is the limiting distribution of a particle:

\medskip
\begin{theorem}\label{theost}
For all finite subsets $I\subset \N$,
\begin{equation*}\label{resEq}
\lim_{N\to \infty} {\cal L}_\st\left((q_i^N(.))_{i\in I}\right)
=Q_\st^{\otimes \vert I\vert} \quad \hbox{weakly.}
\end{equation*}
\end{theorem}

\medskip
Theorems 2, 3 and 4 can be obtained applying classical mean field proof techniques, such as those used in \cite{EK86, S91, G00}. The interested reader is referred to \cite{papermath} for complete proofs of similar results in the case of a much more general system models\footnote{We do not provide the proofs here, but could include them upon editor's request.}.

\subsection{Slotted-Aloha systems - Sequence of type 1}

Consider a type-1 sequence of slotted-Aloha systems as described in the paragraph preceding Theorem 1. In the system with $N$ users, consider a class-$v$ user $i$. When it has a packet in its buffer, it transmits with probability $p_i^N=p_v/N$. The packet arrivals in its buffer are driven by a $\cA$-valued Markov chain $A_i^N$ in stationary regime with distribution $\pi_v$ and whose transition kernel $K_i$ depends on $v$ only. When $A_i^N(t)=a$, a new packet arrives in its buffer with probability $\lambda_{a,v}/N=\lambda_v\times g_{v,a}/N$ (refer to Section \ref{sec1} for the notation).

The system can be represented as a system of interacting particles
as described in Section IV.A. Each user $i$ corresponds to a particle whose state $X_i^N(t)$ at time slot $t$ represents its class $v$, and the length $B_i^N(t)$ of its buffer: $X_i^N(t)=(v,B_i^N(t))$;
i.e. $Y^N_i(t)=B^N_i(t)$. The individual environment of particle $i$ at time slot $t$ is $A_i^N(t)$. Denote by $\nu^N(t)$ the empirical measure of the system at time slot $t$: $\nu^N(t)={1\over N}\sum_{i=1}^N\delta_{X_i^N(t)}$.

Assume that at time slot $t$, the empirical measure is $\nu$. The possible transitions for a user / particle are a packet arrival in the buffer (we index this kind of transition by $b$), and a packet departure (indexed by $d$). Then ${\cal S}=\{b,d\}$. If user (/particle) $i$ is in state $x=(v,k)$, and if its individual environment $A_i^N(t)$ is $a$, the probabilities of transition for the next slot are given as follows. The state becomes $(v,k+1)$ with probability:
$$
F_b^N(x,\nu,a)/N=\lambda_{v,a}/N + o(1/N),
$$
and $(v,k-1)$ with probability:
$$
{F_d^N(x,\nu,a)\over N}={1_{k>0}p_v\over N(1-{p_v\over N})}\prod_{v'}(1-{p_{v'}\over N})^{\beta_{v'}^N\nu_{v'}^{N+}N}+o(1/N),
$$
where $\beta_v^N$ is the proportion of users of class $v$ and $\nu_v^{N+}$ is the proportion of users of class $v$ with non-empty buffers. Denote by $\beta_v$ the
proportion of class-$v$ users at the limit when $N$ grows large.
When $N\to\infty$, the functions $F_b^N$, $F_d^N$ converge
to $F_b$, $F_d$ where:
$$
F_b(x,\nu,a)=\lambda_{v,a},
$$
$$
F_d(x,\nu,a)=1_{k>0}p_v\exp{(-\sum_{v'}\beta_{v'}\nu_{v'}^+p_{v'})}.
$$
One can easily check that Assumptions A1 and A2 are satisfied. Moreover, the limiting averaged transition rates are:
$$
\overline{F}_b(x,\nu)=\lambda_{v},$$
$$
\overline{F}_d(x,\nu)=1_{k>0}p_v\exp{(-\sum_{v'}\beta_{v'}\nu_{v'}^+p_{v'})}.
$$

At time 0, we apply a random and uniformly distributed permutation to the users so that their initial states become i.i.d.. This operation does not change the stability of the system. Finally we scale time and consider $q_i^N(\tau)=X_i^N(\lfloor {N\tau}\rfloor)$. We can apply Theorems 2 and 3, and conclude that when $N$ grows large, the evolutions of the users become
independent. Furthermore, at time $\tau$, if $Q_{(v,k)}(\tau)$ denotes the limiting
probability that a user of class $v$ has $k$ packets in its buffer ($Q_{(v,k)}(\tau)=\lim_{N\to\infty}Pr[q_i^N(\tau)=(v,k)]/\beta_v$), we have:
%\begin{eqnarray}\label{eq:dyn1}
% lefteqn{{\partial Q_{(v,a,k)}\over\partial t}}\\
% &=& \sum_{a'\in \cA}K_v(a',a)Q_{(v,a',k)}(t) -Q_{(v,a,k)}(t)\sum_{a'\in \cA}K_v(a,a')\nonumber\\
%&&+ \lambda_{v,a}\left(  1_{k>0}Q_{(v,a,k-1)}(t)- Q_{(v,a,k)}(t)\right)\nonumber\\
%&&+ p_v\exp{(-\gamma(t))}\left(Q_{(v,a,k+1)}(t)-
% 1_{k>0}Q_{(v,a,k)}(t)\right)\nonumber
% \end{eqnarray}
\begin{align}\label{eq:dyn1}
 {\partial Q_{(v,k)}\over \partial \tau}(\tau)&
 =\lambda_{v}\left(  1_{k>0}Q_{(v,k-1)}(\tau)- Q_{(v,k)}(\tau)\right)\\
+ p_v &\exp{(-\gamma(\tau))}\left(Q_{(v,k+1)}(\tau)-
1_{k>0}Q_{(v,k)}(\tau)\right).\nonumber
\end{align}
with
\begin{equation}\label{eq:dyn2}
\gamma(\tau)=\sum_v \beta_vp_v\nu_v^+(\tau)=\sum_v
\beta_vp_v(1-Q_{(v,0)}(\tau)).
\end{equation}
For a given $v$, equations (\ref{eq:dyn1}) are the Kolmogorov equations corresponding
to the evolution of the number of clients in a queue with Poisson arrivals, exponential service requirements, and time-varying capacity equal to $p_v\exp{(-\gamma(\tau))}$ at time $\tau$, in short to an $M/M_{\tau}/1$ queue. One can also write the evolution of the workload $W_v(\tau)=\sum_k kQ_{(v,k)}(\tau)$ of a queue of class $v$:
\begin{equation}\label{mean}
 {\partial W_{v}\over \partial \tau}(\tau)=\lambda_v -p_ve^{-\gamma(\tau)}(1-Q_{(v,0)(\tau)}).
\end{equation}
Finally, multiplying by $\beta_v$ and summing over $v$, we can
characterize the evolution of the total workload $W(\tau)=\sum_v\beta_vW_v(\tau)$
as:
\begin{equation}\label{summean}
 {\partial \over \partial \tau}W(\tau)=\sum_v\beta_v\lambda_v -\gamma(t)\exp(-\gamma(\tau)).
\end{equation}

\subsection{Slotted-Aloha systems - Sequence of type 2}

Consider now a sequence of slotted-Aloha systems of type 2. Here the Markov chains modulating the arrival processes
evolve at the same rate as the users. The only difference with a sequence of type 1 is then that for any user $i$ of class $v$, we have for all $a\neq a'\in {\cal A}$, $K_i^N(a,a')\approx K_v(a,a')/N$.

Again, the system can be represented as a system of interacting particles, but without individual environments: the state of the Markov chain modulating the arrival process is included in the particle state. Hence we define: $X_i^N(t)=(v,B_i^N(t),A_i^N(t))$; i.e. $Y^N_i(t)=(B_i^N(t),A_i^N(t))$. We have now three types of transitions: arrivals, departures, and changes in the state of the modulating Markov chain. Assume that at time slot $t$, the empirical measure of the system in $\nu$, and consider a particle in state $x=(v,k,a)$. The transition probabilities for the next slot are given as follows. The state becomes $(v,k+1,a)$ with probability:
$$
F_b^N(x,\nu)/N=\lambda_{v,a}/N+o(1/N);
$$
it becomes $(v,k-1,a)$ with probability:
$$
{F_d^N(x,\nu)\over N}={1_{k>0}p_v\over N(1-{p_v\over N})}\prod_{v'}(1-{p_{v'}\over N})^{\beta_{v'}^N\nu_{v'}^{N+}N}+o(1/N),
$$
where $\beta_v^N$ is the proportion of users of class $v$ and $\nu_v^{N+}$ is the proportion of users of class $v$ with non-empty buffers; finally it becomes $(v,k,a')$ with probability:
$$
F_{c(a')}^N(x,\nu)/N=K_v(a,a')/N+o(1/N).
$$
When $N\to\infty$, the functions $F_b^N$, $F_d^N$, and $F_{c(a')}^N$ converge respectively  to $F_b$, $F_d$, and $F_{c(a')}$  where:
$$
F_b(x,\nu)=\lambda_{v,a},\quad F_{c(a')}(x,\nu)=K_v(a,a'),
$$
$$
F_d(x,\nu)=1_{k>0}p_v\exp{(-\sum_{v'}\beta_{v'}\nu_{v'}^+p_{v'})}.
$$
Again one can easily check that Assumptions A1 and A2 are satisfied. Denoting $q_i^N(\tau)=X_i^N(\lfloor N\tau\rfloor)$, we show as previously that if $Q_{(v,k,a)}(\tau)=\lim_{N\to\infty} Pr[q_i^N(\tau)=(v,k,a)]/\beta_v$, then:
\begin{align}\label{eq:dyn1bis}
{{\partial Q_{(v,a,k)}(\tau)\over\partial \tau}}&=\sum_{a'\in \cA}K_v(a',a)Q_{(v,a',k)}(\tau)\\
-&Q_{(v,a,k)}(\tau)\sum_{a'\in \cA}K_v(a,a')\nonumber\\
+ &\lambda_{v,a}\left(  1_{k>0}Q_{(v,a,k-1)}(\tau)- Q_{(v,a,k)}(\tau)\right)\nonumber\\
+p_v\exp&{(-\gamma(t))}\left(Q_{(v,a,k+1)}(\tau)-
1_{k>0}Q_{(v,a,k)}(\tau)\right),\nonumber
\end{align}
where $\gamma(\tau)$ is given by (\ref{eq:dyn2}) with $Q_{(v,k)}(\tau)=\sum_{a}Q_{(v,a,k)}(\tau)$. For a given $v$, equations (\ref{eq:dyn1bis}) are the Kolmogorov equations corresponding to the evolution of the number of clients in a queue with Poisson-modulated arrivals, exponential service requirements, and time-varying capacity equal to $p_v\exp{(-\gamma(\tau))}$ at time $\tau$. In the following, we denote by $M^{K_v}/M_{\tau}/1$ such a queue: the superscript $K_v$ represents the kernel of the process modulating the arrival rates, the subscript $\tau$ means that the capacity is time-varying.

Now for a given class $v$, multiplying (\ref{eq:dyn1bis}) by $k$, and then summing over $a\in \cA$ and $k\ge 0$, one gets (\ref{mean}) where $W_v(\tau)=\sum_k\sum_akQ_{(v,a,k)}(\tau)$; this is due to the fact that we assumed that the Markov chains modulating the arrivals are initially in their stationary regimes, which implies that for any $\tau$, $\sum_k \lambda_{a,v}Q_{(v,a,k)}(\tau)=\lambda_v$. Note finally that (\ref{summean}) is also valid (as a direct consequence of (\ref{mean})).

\section{Asymptotic stability}

\subsection{Stability of the limiting system}

We now investigate the stability of the dynamical system (\ref{eq:dyn1bis})-(\ref{eq:dyn2}) corresponding to a sequence of slotted-Aloha of type 2. Actually, analyzing the stability of (\ref{eq:dyn1bis})-(\ref{eq:dyn2}) is more difficult than analyzing that of the dynamical system (\ref{eq:dyn1})-(\ref{eq:dyn2}) corresponding to a sequence of slotted-Aloha of type 1. As it turns out they have exactly the same stability condition. We let the reader adapt the following analysis to the case of the system (\ref{eq:dyn1})-(\ref{eq:dyn2}).

Assume that $\sum_{v} \beta_v \lambda_v < e^{-1}$. In the following we
denote by $\igam$ and $\sgam$ the unique solutions in $(0,1)$ and in $(1,\infty)$, respectively, of:
\begin{equation}\label{eq:fix2}
\gamma e^{-\gamma}=\lambda:=\sum_v\beta_v\lambda_v.
\end{equation}
Define the function $\xi$ from $[0,\infty)$ to $[0,e^{-1}]$ by
$\xi(x)=x e^{-x}$. Let $\zeta:=\sum_v\beta_vp_v$. The stability of the dynamical system is given
by:

\medskip
\begin{theorem}\label{th:stablim}
(a) Assume that:
\begin{equation}\label{eq:stablim1}
\zeta < \sgam\mbox{ and } \forall v\in \cV, \lambda_v<p_v\exp(-\igam),
\end{equation}
then the dynamical system (\ref{eq:dyn1bis})-(\ref{eq:dyn2}) is globally stable, and $p>\igam$.\\
(b) If for some $ v\in \cV, \lambda_v>p_v\exp(-\igam)$ or if $\zeta <\igam$ then the dynamical system is unstable.\\
(c) If $\zeta >\sgam$, then the system is not globally stable.
\end{theorem}

\medskip
The above theorem states that the stability region of (\ref{eq:dyn1bis})-(\ref{eq:dyn2}) is $\Gamma(1,V)$, where for $b\in [0,1]$, $\Gamma(b,V)$ is the following subset of ${\R}^V_+$:
\begin{equation*}\label{def1}
\{\lambda\in {\R}_+^V: \exists \rho\in [0,1]^V:
\forall v, \lambda_v=p_v\rho_vbe^{-\sum_u\beta_u\rho_up_u}\}.
\end{equation*}
Actually, one can easily prove that $\Gamma(b,V)$ is the stability region of a generalized system obtained from (\ref{eq:dyn1bis})-(\ref{eq:dyn2}) by adding a slot availability probability $b$ to the service rate of class-$v$ users; i.e. this service rate becomes $bp_v\exp(-\gamma(\tau))$. We now provide an alternative representation of $\Gamma(b,V)$. Define $\Lambda(b,V)$ as the subset of $\R_+^V$ whose upper Pareto-boundary is the union of the following surfaces $\partial_v\Lambda(b,V)$:
\begin{align*}
\{\lambda\in {\R}_+^V: \exists &\rho\in \partial_v[0,1]^V :
\forall u,\\
& \lambda_{u}=\rho_{u}p_{u}be^{-\sum_w \beta_w\rho_w p_w}\}.
\end{align*}
In the following, we use the following notation: for all $\xi, \phi\in \mathbb{R}^V$, $\langle\xi,\phi\rangle:=\sum_v\xi_v\phi_v$. We prove that when $\langle\beta,p\rangle =\zeta<1$, then
$\Lambda(b,V)=\Gamma(b,V).$ Let component $v$ of the function $f$ be
$f_v(g)=bg_v\exp(-\langle\beta,g\rangle)$, and let ${\cal P}=[0,p_1]\times\ldots\times [0,p_V]$.
The derivative $df$ of $f$ is $b\exp(-\langle\beta,g\rangle)(I-g \beta^T)$ where $g^T=(g_1,\ldots,g_V)$
and $\beta^T=(\beta_1,\ldots,\beta_V)$.  $(I-g \beta^T)$
 is a rank one matrix with one nonzero eigenvalue, $\langle\beta,g\rangle$
associated with the eigenvector $g$. Since  $\sum_u\beta_u g_u\leq \sum_v\beta_vp_v<1$, the inverse of  $df$ is the positive matrix
$b^{-1}\exp(\langle\beta,g\rangle)(I+(1-\langle\beta,g\rangle)^{-1}g \beta^T)$ by inspection.
$f$ is clearly one-to-one from ${\cal P}$ to the star-like domain $f({\cal P})=\Gamma(b,V)$.
Since $df$ is nonsingular it follows that the image of points $g$ in the interior of ${\cal P}$
are mapped to the interior of $f({\cal P})$ (since $f(g+h)-f(g)$ includes a ball around $g$ by first
order approximation)
and that points $g$ on the boundary of ${\cal P}$ are mapped to the boundary of $f({\cal P})$ (again by
first order approximation). It is also clear that the upper, respectively lower, boundary of ${\cal P}$
is mapped to the upper boundary (the union of the $\partial_v\Lambda(b,V)$),
respectively lower boundary of $f({\cal P})$.
Moreover, since the inverse of $df$ is a positive matrix, it follows that
if $\alpha<\lambda\in \Gamma(b,V)$ then $\alpha\in \Gamma(b,V)$.
It follows that the boundaries of $f({\cal P})$ are Pareto boundaries so $\Lambda(b,V)=\Gamma(b,V).$

\medskip
{\it Proof of Theorem \ref{th:stablim}.} The proof is based on the probabilistic
interpretation of the dynamical system (\ref{eq:dyn1bis})-(\ref{eq:dyn2})
as a collection of $M^K/M_{\tau}/1$ queues: a queue parameterized by $v$ has Markovian
arrivals of intensity $\lambda_v$ and following kernel $K_v$, and it is served
at rate $p_v\exp{(-\gamma(\tau))}$ at time $\tau$.

Now for two probability measures $\sigma,\sigma'$ on $\N\times
{\cal A}$, we write $\sigma\le_{st}\sigma'$ (and say that $\sigma'$ is stochastically greater than $\sigma$) if for all $k\in \N$
and all $a\in {\cal A}$, $\sum_{l=0}^k\sigma_{a,l}\ge
\sum_{l=0}^k\sigma'_{a,l}$. For a collection
$\alpha=(\alpha_v,v\in\cV)$ of probability measures on $\N\times \cA$, we
also define
$\gamma_a^{\alpha}=\sum_v\beta_vp_v(1-\alpha_{v,a,0})$. For two
sets of measures $\alpha,\alpha'$:
\begin{equation}\label{eq:cn1}
\hbox{if }\forall v, \alpha_{v}\le_{st}\alpha'_v,\hbox{ then } \gamma_a^{\alpha}\le\gamma_a^{\alpha'}, \forall a\in {\cal A}.
\end{equation}
Let us now denote by $Q^\alpha(\cdot)$ the set of probability measures
solution of (\ref{eq:dyn1bis})-(\ref{eq:dyn2}) with for all $v$,
$Q_v^\alpha(0)=\alpha_v$. We also define $\gamma^\alpha(\cdot)=\sum_{v\in\cV}\beta_vp_v(1-Q_{(v,0)}^\alpha(\cdot))$, and for all $v$, $W_v^\alpha(\cdot)$ the workload of a queue of type $v$ when the system starts in state $\alpha$.
$Q^0(t)$ is obtained when we
start with an empty system, i.e., $\sum_{a\in\cA}Q_{(v,a,k)}^0(0)=1_{k=0}$ for all
$v$. We have:

\medskip
\begin{lemma}\label{lem:stoc}
If for all $v$, $\alpha_v\le_{st}\alpha'_v$, then
$$
\forall \tau\ge 0, \quad Q^\alpha(\tau)\le_{st} Q^{\alpha'}(\tau).
$$
Furthermore: $\forall \tau,h\ge 0$, $Q^0(\tau)\le_{st}Q^0(\tau+h).$
\end{lemma}

\medskip
\bp The proof of the first statement can be made using (\ref{eq:cn1}) and standard coupling arguments \cite{lindvall}. It suffices to observe that the arrivals are exogenous so we can make the arrival process identical in both copies of the coupled chains. Also note that (\ref{eq:cn1}) implies that the service rates of the queues in the system starting from $\alpha$ remain always greater than those of the queues in the system starting from $\alpha'$. To prove the second statement, observe that for all $v$, $Q^{Q(h)}_{v} \geq_{st} Q^0_{v} (0) = 0$.
Hence by monotonicity, $Q^0 _{v} (\tau+h) \stackrel{\mathcal L}{=} Q^{Q^0(h)}_{v} (\tau)  \geq_{st} Q^0_{v} (\tau)$.
\ep

\medskip
\noindent
\underline{Proof of (a): Stability starting from an empty system.}
Assume that we start from an empty system. Then $\gamma^0(\tau)=0$
and from Lemma \ref{lem:stoc}, $Q_v^0(\tau)$ is stochastically
increasing in time, and $\gamma^0(\tau)$ is a non-decreasing function.
This also implies that $W^0(\tau)$ increases, and then, by (\ref{summean}):
$$
\forall \tau, \lambda\ge \gamma^0(\tau)\exp(-\gamma^0(\tau)).
$$
Remark also that $\gamma^0(\tau)$ converges to some $G$ when $\tau\to\infty$.
From the above equation, we deduce that $G\leq\igam$ since by (\ref{summean}),  $W^0(\tau)$ decreases if
$\gamma^0(\tau)>\igam$. Next $\lambda_v<p_v\exp(-\igam)\le p_v\exp(-G)$ so
the workload $W_v$ is stable as $\gamma^0(\tau)\to G$ and the distribution of queue $v$ is that of an $M^{K_v}/M/1$ queue with service rate $p_v\exp(-G)$. Hence, $G=\sum_v\beta_vp_v \lambda_v/(p_v\exp(-G))$ so $\xi(G)=\lambda$ and $G=\igam$.

Finally, $p > \underline{\gamma}(\lambda)$ is directly deduced from $\lambda_v< p_v\exp(-\underline{\gamma}(\lambda))$.

\medskip
\noindent
\underline{Proof of (a): Arbitrary initial condition.}

We first state a further property of a system starting empty. The result, proved in appendix, says that $\gamma^0(\tau)$ converges {\it rapidly} to $\igam$ when $\tau\to\infty$.

\medskip
\begin{prop}\label{prop:experg}
Define for all $\tau\ge 0$, $f(\tau)=\xi(\igam)-\xi(\gamma^0(\tau))$. Then we have:
\begin{equation}\label{eq:erg}
\int_0^\infty f(u)du < \infty.
\end{equation}
\end{prop}

\medskip
Assume that the initial state is $\alpha=(\alpha_v, v\in \cV)$. By monotonicity, $Q^{\alpha}_{(v,0)}(\tau)\leq Q^{0}_{(v,0)}(\tau)$ for all $v$ and $\tau$. This implies for any $\tau$:
\begin{equation}\label{eq:cn2}
\gamma^{\alpha}(\tau)\geq \gamma^{0}(\tau).
\end{equation}
Note also that $\gamma^{\alpha}(\tau)\le \zeta<\sgam$ (the latter inequality is by assumption). Combining this observation with (\ref{eq:cn2}), we deduce $\xi(\gamma^{\alpha}(\tau)) \ge \xi(\gamma^0(\tau))$, or equivalently
\begin{equation}\label{eq:cn3}
\xi(\gamma^{\alpha}_v(\tau))\ge \xi(\igam) - f(\tau).
\end{equation}
Also by monotonicity: $W^{\alpha}(\tau)\geq W^0(\tau)$. Hence we have:
\begin{align*}
W^0(\infty)-W^{\alpha}(0)&\leq W^\alpha(\infty)-W^{\alpha}(0)\\
=&\int_0^{\infty}{\partial W^{\alpha}\over \partial \tau}(u)du\\
=& \int_0^{\infty}[\xi(\igam)-\xi(\gamma^{\alpha}(u))]du\\
=& \int_0^{\infty}[\xi(\gamma^{\alpha}(u))-\xi(\igam)]^{-}du\\
&-\int_0^{\infty}[\xi(\gamma^{\alpha}(u))-\xi(\igam)]^+du,
\end{align*}
where for any real number $x$, $x^+=\max(0,x)$ and $x^-=-\min(0,x)$. From (\ref{eq:cn3}), we deduce:
\begin{align*}
&W^0(\infty)-W^{\alpha}(0)\\
&\leq \int_0^{\infty}\bigg[f(u)-([\xi(\gamma^{\alpha}(u))-\xi(\igam)]^+)\bigg]du,
\end{align*}
and from (\ref{eq:erg}), conclude that:
$$\int_0^{\infty}[\xi(\gamma^{\alpha}(u))-\xi(\igam)]^+du<\infty.$$
However the derivative of $\gamma^{\alpha}(\cdot)$ is bounded (see
(\ref{eq:dyn1bis})), so we obtain $\xi(\gamma^{\alpha}(\tau))\to
\xi(\igam)$ when $\tau\to \infty$. Finally $\gamma^{\alpha}(\tau)\to \igam$ when $\tau\to \infty$. Now as before, if  $\lambda_v<p_v\exp(-\igam)$, queue $v$ is stable. Summing over $v$
gives $\sgam<\zeta$.

\medskip
\noindent
\underline{Proof of (b):}
Using monotonicity again, to prove the result, we just need to prove instability for a system starting empty. As previously, $\gamma^0(\tau)\to G\leq \igam$ when $\tau\to\infty$. Suppose some queues $\{v\in S^c\}$ are unstable while the rest $\{v\in S\}$ are stable.
 Consequently,
 \begin{eqnarray*}
 G&=&\sum_{v\in S^c}\beta_vp_v+\sum_{v\in S}\beta_vp_v(1-(1-\frac{\lambda_v}{p_v\exp(-G)}))\\
 &=&\zeta'+(\lambda-\lambda')e^{G}
 \end{eqnarray*}
 where $\zeta'=\sum_{v\in S^c}\beta_vp_v$ and $\lambda'=\sum_{v\in S^c}\beta_v\lambda_v$. If there is a $v$ such that $\lambda_v>p_v\exp(-\igam)$, then $G<\igam$ and the workload $W^0$ diverges.
 If  $\zeta<\igam$ then
$\gamma^{0}(\tau)\leq \zeta<\igam$ so it follows from (\ref{summean}) that the workload tends to infinity
and $G<\igam$.

\medskip
\noindent
\underline{Proof of (c):}
We just show here that the dynamical system has two fixed points. We have already shown that, if for all $v$, $\lambda_v<p_v\exp(-\igam)$, and if the system start at 0, then it converges to a fixed point where $\gamma(\tau)=\igam$ and where a queue of subset $v$ has the same distribution as a stationary $M^{K_v}/M/1$ queue of capacity $p_v\exp(-\igam)$. \\
Now the second fixed point is obtained as follows. Assume $\lambda_v<p_v\exp(-\sgam)$ for all $v$. Suppose also that the initial condition for queues of subset $v$ is the stationary distribution of an $M^{K_v}/M/1$ queue with capacity $p_v\exp(-\sgam)$. Then the derivatives in (\ref{eq:dyn1bis}) are all 0 and we have identified a second fixed point.
\ep

\subsection{Stability of the finite system of queues}

To conclude the proof of Theorem \ref{th:stab1}, we need to relate the stability region of the dynamical system (\ref{eq:dyn1bis})-(\ref{eq:dyn2}) to the stability region of the finite system of queues.

\subsubsection{Sufficient ergodicity condition}\label{subsec:suff}

The arrival (resp. transmission) rate of a user $i$ of class $v\in
\cV$ is $\lambda_v/N$ (resp. $p_v/N$). Let
$\lambda^V=(\lambda_1,\ldots ,\lambda_V)$. Then, in this setting,
$\lambda^N+\epsilon\cdot 1^N\in \hat\Lambda^N$ iff
$\lambda^V+\epsilon\cdot 1_V \in \hat\Lambda^N(1,V)$, where $1_V$ is
the $V$-dimensional vector $(1,\cdots,1)$, and for $b\in [0,1]$,
$\hat\Lambda^N(b,V)$ is the subset of $\R^V_+$ whose
Pareto-boundary is the union (over $v$) of the following surfaces:
\begin{eqnarray*}
\{\lambda & :& \exists \rho\in \partial_v[0,1]^V : \forall v',\\
 {}&&\lambda_{v'}={bp_{v'}\rho_{v'}\over 1-\rho_{v'}{p_{v'}\over N}} \prod_u(1-\rho_u{p_u\over N})^{\beta_u^NN}\}.
\end{eqnarray*}
One can easily see that for $N$ large enough, $\hat\Lambda^N(b,V)$ is very close to $\Lambda(b,V)$ (their Hausdorff distance is of order $1/N$). From this we deduce that there exists $N_\epsilon$, such that for all $N>N_\epsilon$ , $\lambda^V+\epsilon\cdot 1_V \in \hat\Lambda^N(1,V)$. Define $\hat\Gamma^N(b,V)$ as:
\begin{eqnarray*}\label{def1}
\{\lambda\in {\R}_+^V &:& \exists \rho\in [0,1]^V: \forall v,\\
{}&& \lambda_v={bp_v\rho_v\over 1-\rho_v{p_v\over N}}\prod_u(1-\rho_u{p_u\over N})^{\beta_u^NN}\}.
\end{eqnarray*}
We can prove (as done after Theorem 5) that when $\sum_v\beta_vp_v<1$, $\hat\Lambda^N(b,V)=\hat\Gamma^N(b,V)$.

We now consider systems built from our original systems but such that each slot is available for transmission with probability $b$, i.i.d. over slots. We show the following result by induction on $V$, and deduce Theorem 1 (a) applying it for $b=1$.

``If there exists an $\epsilon>0$ small enough, such that for $N$ sufficiently large, $\lambda^V+\epsilon\cdot 1_V \in \hat{\Lambda}^N(b,V)$,
then the system with $N$ queues is stable.
Furthermore in such a case, the stationary distributions $\pi_{st}^N$ of such systems
constitute a tight family of probability measures.''

Let us first prove the result when $V=1$. In such case, all the queues are similar, and the system is then homogenous. We have $\hat{\Lambda}^N(b,1)=\Lambda^N(b,1)$ and the system is stable iff $\lambda_v < p_v b(1-p_v/N)^{N-1}$ by \cite{Szpan94}. Now assume that $\lambda_v < p_v b(1-p_v/N)^{N-1}-\epsilon$. Consider a particular queue: at any time, its distribution is stochastically bounded by the distribution we would obtain assuming that all the other queues are saturated. In the latter system, the stationary distribution of the queue considered is that of a Markovian queue of load $\lambda_v/(p_v b(1-p_v/N)^{N-1})<1-\alpha\epsilon$ for some $\alpha>0$. Tightness follows.\\

Now let us assume that the result is true when $\vert {\cal V}\vert\le V$, and let us prove it when $\vert {\cal V}\vert = V+1$. Assume that for $N$ large enough, $\lambda^{V+1} +\epsilon\cdot 1_{V+1} \in \hat{\Lambda}^N(b,V+1)$. Denote by $\lambda^{V+1,v}$ the $V$-dimensional vector built from $\lambda^{V+1}$ where the $v$-th component has been removed. Since $\hat\Lambda^N(b,V+1)=\hat\Gamma^N(b,V+1)$, there exists $v$ such that: for $N$ large enough,
\begin{equation}\label{eq21}
\lambda^{V+1,v}+\epsilon\cdot 1_V \in \hat\Lambda^N\bigg(b(1-{p_v\over N})^{\beta_v^NN},V\bigg).
\end{equation}
Consider the stochastically dominant system where all queues of class different than $v$ see saturated queues of class $v$. For the latter sub-system, in view of (\ref{eq21}), we can apply the induction result. We conclude that for $N$ large enough, the dominant system without queues of class $v$ is stable, and that the family of the corresponding stationary distributions $\pi_{st}^{N,v}$ is tight. \\
From Theorem 5 applied to the dominant systems without queues of class $v$, we know that the corresponding limiting system is globally stable. We can then apply Theorem 4 to these systems to characterize the average proportion of slots left idle by the queues of class different than $v$: when $N\to\infty$, this proportion tends to $\exp{(-\sigma})$ where $\sigma$ is the lower solution of $\sigma e^{-\sigma}.be^{-\beta_vp_v}=\sum_{u\neq v}\beta_u\lambda_u$. Now consider a queue of class $v$ in the dominant system. Denote by $(S_t^N, t\ge 0)$ its service process. We can make this process stationary ergodic just assuming that initially the system without queues of class $v$ is in stationary regime. The service rate of a queue of class $v$ converges to $bp_v\exp(-\sigma)\exp(-\beta_vp_v)$ when $N\to\infty$. Hence when $N$ is large enough, we have:
$$
E[S_t^N] \ge bp_v\exp(-\sigma)\exp(-\beta_vp_v)-\epsilon/4.
$$
Now since $\lambda^{V+1}+\epsilon\cdot 1_{V+1} \in \hat\Lambda^N(b,V+1)$, we have for $N$ large enough:
$$
\lambda_v< bp_v\exp(-\sigma)\exp(-\beta_vp_v)-\epsilon/2\le E[S_t^N] -\epsilon/4.
$$
We deduce that in the dominant system, the queue of class $v$ are stable for $N$ large enough, and that their stationary distributions are tight. We conclude the proof noting that the original systems are stochastically dominated by systems that are stable for $N$ large enough, and such that their stationary distributions are tight.

\subsubsection{Necessary ergodicity condition}

We now prove that there exists an integer $N_\epsilon$ such that
for all $N \geq N_\epsilon$, the system is unstable if $\lambda^N - \epsilon .  1^N \notin \hat \Lambda^N $ or, equivalently, if
\begin{equation} \label{eq:hypo1}
\lambda^V - \epsilon .  1_V \notin  \hat
\Lambda^N(1,V),
\end{equation}
where $\hat \Lambda^N(1,V)$ is
defined in the previous paragraph. The system is monotone with
respect to the arrival process: if we remove some incoming
packets, the buffers cannot increase. Hence it is sufficient to
prove that a modified system obtained from an independent thinning
of the arrival process of all users is unstable.

We first perform this suitable thinning. By assumption, there
exists $t^N \in (0,1)$ such that $t^N \lambda^V \in \hat
\Lambda^N(1,V)$. Then, there exists a class $v$ such that $t^N
\lambda^V \in \partial_v \hat \Lambda^N(1,V)$. Up to extracting a
subsequence, we may assume that this class $v$ does not depend on
$N$, for $N$ large enough. Note also that the convergence of $\hat
\Lambda^N(1,V)$ to $\Lambda(1,V)$ implies that $t^N$ converges to
some $t^* \in (0,1)$ such that $t^* \lambda^V \in
\partial_v \Lambda(1,V)$. We assume for
simplicity that there exists a unique class $v$ like that (the
proof generalizes easily by performing a non-homogeneous
thinning). Then, from Assumption (\ref{eq:hypo1}), we can choose
$\eta>0$ small enough such that $t := t^* + \eta \in (t^* , 1)$
satisfies for some $\epsilon'>0$ and all $N$ large enough,
$$
t \lambda^{V,v} + \epsilon' 1_{V-1} \in  \hat \Lambda^N((1 - \frac{
p_v}{ N})^{\beta_v^N N},V-1),
$$
and
$$
t \lambda^{V,v} > p_v \exp ( - \sigma^t) \exp( - \beta_v p_v) +
\epsilon',
$$
where $\sigma^t$ is the smallest
solution of $\sigma \exp (\sigma) \exp (- \beta_v p_v) = \sum_{u
\ne v} \beta_u t \lambda_u$. We now perform the thinning of the arrival processes: Up to replacing $\lambda^V$ by $t\lambda^V$, we can assume directly that for all $N$ large enough,
$\lambda^{V,v} + \epsilon' 1^{v} \in \hat \Lambda^N((1 - \frac{ p_v}{
N})^{\beta_v^N N},V-1)$ and $\lambda_v >   p_v \exp ( - \sigma)
\exp( - \beta_v p_v) + \epsilon'$. Now, we define the stopping
time
\begin{align*}
T^N_v &= \inf \{ t \geq 1 : \hbox{there exists a class-$v$ user
whose} \\
& \hbox{buffer is smaller than } 1+ \log t \}.
\end{align*}

If we prove that for an arbitrary initial condition, $\PP (T ^N _v
= \infty)>0$ then the system is transient.  As in the previous
paragraph, we consider the dominant system where all users of
class different than $v$ see saturated class-$v$ users. Recall
that, by construction, in this dominant system, the buffers of
class-$v$ users are independent of the buffers of users of class
different than $v$. Note also that on the event $\{T^N_v \geq t\}$
all class-$v$ users are saturated on $[0,t]$, hence on this time
interval the dominant system and the original system couple. From
the strong Markov property, it implies that $\PP (T^N_v = \infty)$
is equal to the probability that for all $t \geq 1$, the buffer of
all class-$v$ users in the dominant system is larger than $1 +
\log t$. Now, we may argue as in the \S \ref{subsec:suff}: for $N$
large enough, the dominant system restricted to users of class
different than $v$ is stable, and the asymptotic proportion of
slots left idle by the queues of class different than $v$ tends to
$\exp(-\sigma)$ as $N$ goes to infinity. Hence, for class-$v$
users, in the dominant system, the service rate converges to $p_v
\exp(-\sigma) \exp( - \beta_v p_v) < \lambda_v - \epsilon'$. In
other words, the size of the buffers of class-$v$ users has a
positive drift and by a routine computation, it implies easily
that there exists $N_\epsilon$ such that for all $N\geq
N_\epsilon$, $\PP (T ^N _v = \infty)>0$.

\section{Extensions: The case of CSMA protocols}

The results of the previous sections can be extended to the case of more efficient random multiple access methods. For example, in CSMA systems, a user senses the channel before transmitting. If the channel is busy, the user remains silent, and when the channel is idle, the user can attempt to transmit packets. This allows users to transmit for a large number of consecutive slots without being interrupted, and thus significantly increases the system efficiency. For example, in the case of slotted-Aloha, it can be easily seen that when the channel is shared by $N$ users with similar characteristics, i.e., same arrival rate $\lambda/N$ and same transmission probability, then the maximum amount of traffic $\lambda$ that the system can support is $e^{-1}$. With CSMA, this amount can be made arbitrarily close to 1 (as we would obtain using a perfect centralized multi-access scheme) just letting the channel holding time grow large.

\subsection{Model}

The model is similar than the one used for slotted-Aloha, except that when a user attempts to use the channel, it keeps transmitting during $\sigma$ consecutive slots. In the current IEEE802.11g standard \cite{802.11g}, for data packets, we have roughly $\sigma\approx 10$ slots. There are $N$ users sharing the resource, and at the beginning of each slot, user $i$ transmits with probability $p_i$ if it senses the channel idle. Packets whose transmission take $\sigma$ slots arrive according to a stationary ergodic Markovian process of intensity $\lambda_i$ in the buffer of user $i$.

\begin{remark}[Heterogeneous systems] The analysis can be generalized to the case where users transmit at different rates, i.e., when user $i$ keeps the channel for $\sigma_i$ slots. To keep the formulas simple we just assume that all users transmit packets of the same sizes and at the same rate, for all $i$, $\sigma_i=\sigma$. Another possible generalization is to allow the collision to be shorter than the packet successful transmissions. This can be useful when one wants to model RTS/CTS signaling scheme in IEEE802.11-based systems.
\end{remark}

\subsection{Approximate stability region}

In CSMA systems, users are synchronized in the sense that they observe the same periods where the channel is busy. The analysis of these systems can then be conducted as that of slotted-Aloha: it suffices to analyze the system at the instants corresponding to the beginning of idle slots or to the beginning of transmissions. The approximate stability region is then constructed as follows.

For $\rho=(\rho_1,\ldots,\rho_N)\in\R_+^N$, define $\gamma_i(\rho)$:
$$
\gamma_i(\rho,\sigma)={P_i\over \sigma(\sum_jP_j+C)+E},
$$
where
$$
\left\{
\begin{array}{l}
P_i=\rho_ip_i\prod_{j\neq i}(1-\rho_jp_j),\\
E=\prod_k(1-\rho_kp_k),\\
C=1-E-\sum_jP_j.
\end{array}
\right.
$$
The approximate stability region $\hat\Lambda^N_\sigma$ is the set of points lying below one of the boundaries $\partial_j\hat\Lambda^N_l$ defined by:
$$
\partial_j\hat\Lambda^N_\sigma =\bigg\{\lambda: \exists \rho\in \partial_j[0,1]^N, \forall i,  \lambda_i=\gamma_i(\rho,\sigma)\bigg\}.
$$

Under the assumptions of Theorem \ref{th:stab1} (see the paragraph above Theorem \ref{th:stab1}), we can show that $\hat\Lambda^N_\sigma$ tends to the actual stability region when $N$ grows large. As in the case of unit packet duration, we can show using theoretical arguments and numerical experiments that the approximation is extremely accurate. For example, the notion of $k$-homogenous directions can be easily extended, and $\hat\Lambda_\sigma^N$ is exact in those directions.

\section{Conclusion}

We have provided a very accurate approximate stability region for
classical slotted-Aloha systems. This approximate region has been
shown to be exact when the number of users becomes large, but is
also extremely accurate for small systems. The analysis has been
generalized to the case of CSMA systems.

In this paper, we have considered network scenarios where all
users share a common channel, and that only a single user can
transmit successfully at a time. An important question that
remains to be investigated is the case where users do not
interfere with all other users, i.e., several users can
simultaneously transmit successfully. For example, a popular model
in the literature consists in modeling user interaction by an
interference graph. In such network scenarios, how efficient are
non-adaptive CSMA protocols? We have provided a preliminary analysis of such network scenarios in \cite{sig08}, but without presenting complete proofs and without being able to deduce results characterizing the efficiency of CSMA protocols.

\bibliographystyle{abbrv}

%\bibliography{Bibfull,wlannet}

\appendix

\noindent
{\it Proof of Proposition 1.}

Consider systems obtained from the original systems but where each slot is available for transmission with probability $b$, i.i.d. over slots. We denote by $\Lambda^N(b)$ the corresponding stability region (of course it depends on the arrivals rates, transmission probabilities and modulating Markov for the arrival processes). We prove the result using Szpankowski recursive expression for $\Lambda^N(b)$. Let us show by induction on $k$ that the following result holds:\\
"For $k$-homogeneous systems with slots available with probability $b$, and such that $1_{k+1\le N}\alpha_{k+1}(1-p_{k+1})/p_{k+1}\le  \alpha_{1}(1-p_{1})/p_1$ and $\alpha_l=0$ for $l>k+1$, we have:
$$
s^\star=\hat{s}^\star={b\prod_{i=1}^k(1-p_i)\over {1-p_1\over p_1}\alpha_1+\alpha_{k+1}}."
$$
For $k=1$, the result follows from the stability analysis of systems with two queues only. Assume that the result holds for all $l<k$. Consider a $k$-homogeneous system. From \cite{Szpan94}, we know that the system is stable if and only if there exists $i\in \{1,\ldots,k+1\}$ such that:
\begin{equation}\label{eq:szp1}
\lambda^i\in \Lambda^k(b(1-p_i)),
\end{equation}
\begin{equation}\label{eq:szp2}
\lambda_i < bp_i\sum_{{\cal L}\subset [k+1]\setminus\{i\}} \pi_{\cal L}\prod_{j\in {\cal L}}(1-p_j),
\end{equation}
where $\lambda^i=(\lambda_1,\ldots,\lambda_{i-1},\lambda_{i+1},\ldots,\lambda_{k+1})$, $[k+1]=\{1,\ldots,k+1\}$, and $\pi_{\cal L}$ denotes the stationary probability that the buffers from set ${\cal L}$ are not empty in a system where user $i$ has been removed and $b$ has been replaced by $b(1-p_i)$. Note that (\ref{eq:szp1}) ensures that these probabilities exist.

Remark that when removing user $i$, the remaining system is $(k-1)$-homogeneous. By induction, if $\lambda=s\alpha$, we deduce that for any $i\neq k+1$, condition (\ref{eq:szp1}) is equivalent to:
\begin{equation}
s< s^\star=\hat{s}^\star={b\prod_{i=1}^k(1-p_i)\over {1-p_1\over p_1}\alpha_1+\alpha_{k+1}}.
\end{equation}
When the latter condition is satisfied, it is easy to show that buffer $i$ in the original system is stable, i.e., that (\ref{eq:szp2}) holds. Indeed consider the stochastically dominant system where users $j\neq i,k+1$ always transmit with probability $p_j$. Then the stability condition of buffers $i$ and $k+1$ is that of a two-buffer system with slot-availability probability equal to $c=b\prod_{j\neq i,k+1}(1-p_j)$. The latter system is stable if and only if:
$$
\lambda_i<cp_i(1-{\lambda_{k+1}p_{k+1}\over c(1-p_i)}),
$$
which is equivalent to:
$$
s<s'= {cp_i\over \alpha_i}(1-{\lambda_{k+1}p_{k+1}\over c(1-p_i)}).
$$
One can verify that $s'\ge s^\star$.

One can also show that conditions (\ref{eq:szp1})-(\ref{eq:szp2}) with $i=k+1$ implies stronger restrictions on $s$ than similar conditions for $i\le k$, which concludes the proof.

\medskip
\noindent
{\it Proof of Proposition \ref{prop:experg}.}

To prove (\ref{eq:erg}), we compare the system with another system that starts empty too, and whose evolution is characterized by (\ref{eq:dyn1bis}) where  $\gamma^0(\tau)$ is replaced by $\igam$. We denote by $Q^{0,e}_{(v,a,k)}(\tau)$ the solution of this new system, and define $\gamma^{0,e}(\tau)=\sum_v\beta_vp_v(1-Q_{(v,0)}^{0,e}(\tau))$. We also denote by $W^{0,e}(\tau)$ its total workload at time $\tau$. The new system is equivalent to a system of $V$ independent queues with Poisson modulated arrivals and constant capacities (equal to $p_v\exp (-\igam)$ for type-$v$ queue). Note that the service rates of the queues in the new system are smaller than those in the original system. We deduce: for all $\tau \ge 0$,
$$
\xi(\gamma^{0,e}(\tau)) \ge \xi(\gamma^0(\tau)).
$$
Also remark that the original and the new systems have the same stationary behavior, which implies: $W^0(\infty)=W^{0,e}(\infty)$. Then:
\begin{align*}
0&=W^0(\infty)-W^{0,e}(\infty)\\
& = \int_0^\infty \bigg[ \xi(\gamma^0(u))-\gamma^{0,e}(u)\exp(-\igam)\bigg]du.
\end{align*}
Hence we have:
\begin{align*}
\int_0^\infty f(u)du &= e^{-\igam} \int_0^\infty \bigg(\gamma^{0,e}(u)-\igam\bigg)du\\
=e^{-\igam} &\int_0^\infty \sum_vp_v\beta_v\bigg(Q_{(v,0)}^{0,e}(\infty)-Q_{(v,0)}^{0,e}(u)\bigg)du.
\end{align*}
To prove (\ref{eq:erg}), it suffices to show that for all $v\in \cV$, $Q_{(v,0)}^{0,e}(\tau)$ converges exponentially fast to $Q_{(v,0)}^{0,e}(\infty)$ when $\tau\to\infty$. As mentioned previously, $Q_{(v,k)}^{0,e}(\tau)$ represents the probability that a queue, initially empty, with Poisson modulated arrivals, exponential service requirements and constant capacity $p_v\exp(-\igam)$. Thus to prove (\ref{eq:erg}), we just need to show that a queue $M^K/M/1$ with Poisson modulated arrivals is exponentially ergodic\footnote{An ergodic and stationary Markov process is exponentially ergodic, if its distribution converges exponentially fast to the its stationary distribution.} under the usual stability condition. This is what we prove next.

\medskip
{\it Exponential ergodicity of queues with Poisson modulated arrivals.} The proof of the exponential ergodicity of queues with Poisson modulated arrivals can be done classically, showing that the spectral gap of the corresponding Markov process is positive. We give the proof for completeness.

Consider a queue of capacity $1/\mu$. Clients arrive according to a Poisson modulated process. The service requirements are i.i.d. exponentially distributed with mean 1. The arrivals are modulated by a $\cA$-valued Markov process $(A(\tau),\tau\ge 0)$ whose transition kernel is $K$, and stationary distribution $\eta$. $\cA$ is a finite set. When $A(\tau)=a$, clients arrive at rate $\lambda_a$. Assume that the queue is ergodic, i.e., $\sum_{a\in\cA}\eta(a)\lambda_a \times \mu <1$. Denote by $B(\tau)$ the number of clients in the queue at time $\tau$. Let $R$ denote the kernel of the Markov process $X=((B(\tau),A(\tau)), \tau\ge 0)$. We have: for all $k\in \N$, $a\in \cA$,
\begin{align*}
&R((k,a);(k-1,a))=1_{k>0}\mu,\\
&R((k,a);(k+1,a))=\lambda_a, \\
&R((k,a);(k,b))=K(a,b).
\end{align*}
Let $\pi(z)$ denote the stationary probability to be in state $z=(k,a)\in \N\times \cA$. Following \cite{liggett}, to prove exponential ergodicity, we just need to show that the spectral gap of $R$ is strictly positive. The gap is defined by:
$$
{\rm{gap}}(R)=\inf_{g\in \cF}\bigg[\frac{1}{2}\sum_{w,z\in \N\times \cA}(g(w)-g(z))^2R(z;w)\pi(z)\bigg],
$$
where $\cF$ is the set of measurable functions such that $\sum_z g(z)\pi(z)=0$, and $\sum_z g^2(z)\pi(z)=1$. The following lemma states the exponential ergodicity of the queue.

\medskip
\noindent
\begin{lemma} 
$$
{\rm{gap}}(R)>0.
$$
\end{lemma}

\medskip
{\it Proof.} We can first show that there exist two strictly positive constants $c_1,c_2$ such that: for all $(k,a)\in \N\times \cA$,
\begin{equation}\label{eq:chee}
c_1\pi_0(k,a) \le \pi(k,a)\le c_2\pi_0(k,a),
\end{equation}
where $\pi_0(k,a)=\eta(a)(1-\rho)\rho^k$ for some $\rho <1$. Actually, this result can be obtained  using one of the classical methods to derive the tail of the stationary distribution of a queue with modulated arrivals, for example refer to Theorem 2.4 in \cite{beck}.

%After uniformization this process $X$ is covered by the models in \cite{beck}.  Using Theorem 2.4 there
%one gets that $\pi(i,a)$ decays like $\exp(-\theta i)$ where $\exp(\theta i)\hat{h}(a)$ is a harmonic function for $X^{\infty}$.
%Since $\exp(\theta i)\hat{h}(a)$ is harmonic $\theta$ and the $\hat{h}(a)$ satisfy
%\begin{eqnarray*}
%\lefteqn{e^{\theta i}\hat{h}(a)(\sum_bK_{ab}+\mu+a)}\\
%&=&e^{\theta i}\sum_bK_{ab}\hat{h}(b)+\hat{h}(a)\mu e^{\theta(i-1)}+\hat{h}(a)ae^{\theta(i+1)}
%\end{eqnarray*}
%or
%\begin{eqnarray}\label{solvit}
%\hat{h}(a)(\sum_bK_{ab}-\mu(e^{-\theta}-1))-a(e^{\theta}-1))&=&\sum_bK_{ab}\hat{h}(b).
%\end{eqnarray}
%Multiply both sides by $\pi(a)$ and sum over $a$ to get
%\begin{eqnarray*}
%&&\sum_a\pi(a)\hat{h}(a)(1-\mu(e^{-\theta}-1)-\sum_a\pi(a)a\hat{h}(a)(e^{\theta}-1))\\
%&=&\sum_a\pi(a)\hat{h}(a).
%\end{eqnarray*}
%Let $\overline{a}=\sum_a\pi(a)a\hat{h}(a)/\sum_a\pi(a)\hat{h}(a)$. Solving we get $\exp(\theta)=\mu/\overline{a}$.
%Substituting this solution back into (\ref{solvit}) gives $|\mathcal{A}|$ equations in $|\mathcal{A}|$ unknowns $\hat{h}(a)$.
%Letting $\rho=\exp(-\theta)$ define $\pi_0(i,a)=(1-\rho)\rho^i\cdot \pi(a)$. We conclude
%we can  pick  constants $c_1,c_2>0$ so that $c_1\pi_0(i,a)\leq \pi(i,a)\leq c_2 \pi_0(i,a)$ for all $(i,a)$.

Note that $\pi_0(k,a)$ is the steady state distribution of a Markov process $X_0$ with two independent components: $X_0=(B_\rho(\tau),A(\tau),\tau\ge 0)$. $B_\rho$ is the Markov process representing the number of clients in an $M/M/1$ queue with load $\rho$, and $A$ and $B_\rho$ are independent. Denote by $R_0$ the transition kernel of $X_0$. 

From (\ref{eq:chee}), one can easily deduce that there exists a constant $c>0$ such that $k_R \ge c k_{R_0}$, where $k_R$ (resp. $k_{R_0}$) is the Cheeger constant of $X$ (resp. $X_0$), see \cite{sokal}. For example, $k_R$ is defined by:
\begin{eqnarray*}
k_R&=&\inf_{H\in\N\times \cA;0<\pi(H)<1}k_R(H)\mbox{ where }\\
k_R(H)&=&\frac{\sum_{(k,a)\in H}\pi(k,a)R((k,a);H^c)}{\pi(H)\pi(H^c)}.
\end{eqnarray*}
The Cheeger constant and the spectral gap are related. Actually thanks to Theorems 2.1 and 2.3 in \cite{sokal}, there exists a constant $C>0$ such that: 
$$
C\times k_R^2\le {\rm{gap}}(R)\le k_R.
$$
The same inequalities (with a different constant $C$) holds for $R_0$. Now observe that ${\rm{gap}}(R_0)>0$. This is due to the fact that by  Theorem 2.6 in \cite{liggett}, ${\rm{gap}}(R_0)$ is the minimum of the spectral gap of $B_\tau$ and that of $A$. Both are strictly positive ($B_\tau$ is a birth-death process, see Corollary 3.8 in \cite{liggett}; $A$ can take a finite number of values). We conclude:
$$
0<{\rm{gap}}(R_0)\le k_{R_0}\le c^{-1}k_R\le c^{-1} \sqrt{C^{-1}{\rm{gap}}(R)}.
$$
Hence ${\rm{gap}}(R)>0$.

%{\it Proof of (ii).} Knowing that the spectral gap of $X$ is strictly positive, we can apply the same analysis as in \cite{liggett} Section 2 (more precisely, apply the $L^2$ analysis of p. 409 in \cite{liggett} to the function $f$ such that for all $(l,a)\in \N\times\cA$, $f(l,a)=1_{k=l}$), and get $(ii)$.  

%.  Moreover the spectral gap of a birth and death process is positive by . The Markov process $A$ has a finite state so it will also have a positive spectral gap. Hence $gap(R_0)>0$.
%We conclude $0<gap(R_0)<k_{R_0}<\zeta k_R$.  Finally by Theorem 2.3 in \cite{sokal} we have
%$gap(R)> \kappa k_R^2/(8M)$ where $M$ is the maximum jump rate and $\kappa$ is a constant given in \cite{sokal}.

%We conclude the process $X$ has a positive spectral gap so we have exponential $L^2$ convergence as in Section 2 in \cite{liggett}.
%Let $f$  there be the indicator function $\chi_j\{\cdot\}$ and $S(t)$ there is the transition kernel of $X(t)$.
%\begin{eqnarray*}
%& &|P(B(t)=j|A(0)=a,B(0)=i)-\pi(j)|^2\pi(i,a)\\
%&=&|S(t)f(i,a)-E_{\pi}f(X(0))|^2\pi(i,a)\\
%&\leq& \sum_{i,a}|S(t)f(i,a)-E_{\pi}f(X(0))|^2\pi(i,a)\\
%&\leq&\exp(-gap(\Lambda)t)\sum_{i,a}|f(i,a)-E_{\pi}f(X(0))|^2\pi(i,a)\\
%&\leq&\exp(-gap(\Lambda)t)( (1-\pi(j))^2\pi(j)+\pi(j)^2(1-\pi(j)))
%\end{eqnarray*}

\end{document}